\documentclass[12pt]{article}
\usepackage{amsmath}
\usepackage{amsfonts}
\usepackage{amssymb} 
\usepackage{graphicx,psfrag,epsf}
\usepackage{enumerate}
\usepackage{enumitem}
\usepackage{natbib}
\usepackage{url}
\usepackage{booktabs}
\usepackage{float}

\usepackage{amsthm}
\theoremstyle{definition}
\newtheorem{definition}{Definition}

\newcommand{\blind}{1}

\begin{document}

\def\spacingset#1{\renewcommand{\baselinestretch}%
{#1}\small\normalsize} \spacingset{1}


\if1\blind
{
  \title{\bf Topological Effective Connectivity Modeling in Brain Networks}
  \author{
    Anass El-Yaagoubi \hspace{.2cm}\\
    Statistics Program, King Abdullah University of Science and Technology
    \and
    Moo K. Chung \\
    Department of Statistics, University of Wisconsin-Madison
    \and
    Hernando Ombao \\
    Statistics Program, King Abdullah University of Science and Technology
  }
  \maketitle
} \fi

\if0\blind
{
  \bigskip
  \bigskip
  \bigskip
  \begin{center}
    {\LARGE\bf Topological Effective Connectivity Modeling in Brain Networks}
\end{center}
  \medskip
} \fi

\bigskip
\begin{abstract}
    Characterizing directed information flow in brain networks is difficult because neural circuits are full of recurrent feedback loops. Many existing tools for directed dependence assume a directed acyclic graph (DAG) structure to resolve directional ambiguity, and therefore cannot represent these loops. We present a nonparametric, information-theoretic framework that addresses this by coupling the discrete Hodge decomposition with lead-lag mutual information, splitting the resulting edge flow into three orthogonal components: a gradient term capturing hierarchical, feed-forward relationships; a curl term isolating triangle-level feedback loops; and a harmonic term capturing cyclic flow around topological holes. This separation makes it possible to disentangle feed-forward drive from recurrent circulation, which conventional measures conflate. We further develop a permutation-based hypothesis-testing layer that identifies nodes and triangular motifs whose information-flow signatures change significantly between conditions. We validate the framework on simulations with known ground-truth structure and apply it to local field potential recordings from a rodent model of focal ischemic stroke. In three of four animals, we find a post-stroke shift toward hierarchical, source-driven propagation at the expense of recurrent feedback, while the fourth shows no significant change.
\end{abstract}

\noindent%
{\it Keywords:} Topological Data Analysis, Higher-Order Networks, Hodge Decomposition, Effective Connectivity, Ischemic Stroke, Local Field Potentials.
\vfill

\newpage
\spacingset{1.45}

\section{Introduction}
\label{sec:intro}

The study of dependence structures within the brain is of fundamental importance in neuroscience, providing critical insights into how information is processed, integrated, and regulated \citep{BRAIN_NETWORKS_1, BRAIN_NETWORKS_2}. By examining brain connectivity we can discern interaction patterns among regions, deepen our understanding of functional integration, and clarify how neurological disorders alter these networks. Traditional approaches rely on graph-theoretical measures such as node centrality, clustering coefficients, and modularity \citep{GRAPH_MODELING_COMPLEX_NETWORK, GRAPH_MODELING_HUMAN_BRAIN, GRAPH_MODELING_BRAIN_CONNECTIVITY}. While these methods have provided valuable insights into pairwise interactions, they are limited in that they fail to capture the higher-order, multi-region dependencies that are essential for understanding complex neural systems.

In recent years there has been growing interest in studying brain connectivity from a higher-order perspective that accounts for interactions beyond simple pairwise associations \citep{HOI_1,HOI_2,HOI_3}. The inherent complexity of brain structure, combined with advances in topological data analysis (TDA) and information-theoretic methods, has enabled powerful new frameworks for investigating these intricate patterns \citep{TDA_BRAIN, TDA_BRAIN_ARTERY, IT_1, IT_2}.

Most topological analyses of multivariate signals have relied on symmetric dependence measures such as correlation, coherence, partial correlation, or partial coherence to build functional brain-connectivity networks \citep{TDA_BRAIN_PROS_AND_CONS, METHODS_BCA}. Because these metrics ignore direction, they capture only undirected pairwise associations. Recently, however, topological methods have been extended to effective connectivity, which models directed statistical dependence among brain regions \citep{TDA_ORIENTED_1, TDA_ORIENTED_2, TDA_ORIENTED_3, TDA_ORIENTED_4}.

The directed dependence we analyze is what the brain-connectivity literature calls effective connectivity: directed statistical dependence between recorded signals, in which lead-lag asymmetry is taken to indicate that the past of one channel carries predictive information about the present of another. We do not use the term in the interventional sense of the Pearl \citep{PEARL_CAUSALITY} or Rubin \citep{RUBIN_1974} frameworks, which require either an explicit intervention or a structural model identified under such an intervention. We do not attempt causal discovery, and we do not claim to identify interventional effects from observational LFP recordings. What we do is take a directed dependence network, however estimated, and decompose it into structurally distinct components, then test which components change between experimental conditions. The experimental stroke in our application is itself an intervention at the system level, perturbing the brain and allowing us to compare the resulting dependence networks before and after; whether the observed changes reflect direct or mediated mechanisms is a separate question.

Directed analyses of time series often assume that the underlying structure is a directed acyclic graph (DAG) \citep{CI_ASSUMPTIONS_1, CI_ASSUMPTIONS_2}. This assumption is incompatible with effective brain connectivity, where cortical and subcortical regions continually exchange signals and produce recurrent feedback loops. Such cycles violate acyclicity, blur temporal precedence (cause and effect can occur within the same sampling interval), and allow latent common drivers to generate spurious links. All of these undermine standard DAG-based causal-discovery algorithms. Faced with feedback, these algorithms must either sever cycles, thereby losing vital physiological information, or return partially oriented, ambiguous graphs. Crucially, feedback loops exist in the brain whether or not a given algorithm chooses to model them; our approach therefore takes the directed dependence network as given and asks how to disentangle its overlapping cyclic and acyclic structure.

To this end we propose Topological Effective Connectivity Modeling (TECM), a framework for analyzing directed dependence in networks where feedback and recurrence cannot be assumed away. Rather than running a causal-discovery algorithm that forces the network to be acyclic, TECM begins with a flexible prior structure on the set of possible directed edges, which can range from spatial adjacency (as in our LFP example) to anatomically informed connectivity priors (e.g., from diffusion imaging) or fully connected networks where the testing layer suppresses spurious edges. Given this skeleton, TECM estimates edge-wise influence using lead-lag mutual information, with the $k$-nearest-neighbor estimator of \citet{KRASKOV_MI}. The result is a weighted, directed graph that represents putative directed dependence. The discrete Hodge decomposition \citep{HODGE_RANK, HODGE_GRAPHS} is then applied to this graph, partitioning the total information flow into three orthogonal components: a gradient term that captures feed-forward, DAG-like structure and induces a ranking of nodes from sources to sinks; a curl term that captures feedback flow circulating around individual triangles; and a harmonic term that captures cyclic flow around topological holes that cannot be reduced to triangle-level circulation. The spatial extent of the curl and harmonic components is therefore determined by the chosen triangulation: in our application, triangles are restricted to spatially adjacent electrodes, so curl captures short-range feedback among neighboring sites and harmonic captures longer-range cycles around topological holes in the local connectivity. Building on this decomposition, TECM adds a hypothesis-testing layer that identifies which nodes and triangular circuit motifs exhibit statistically significant changes in information flow between conditions, while properly accounting for multiple comparisons \citep{BENJAMINI_HOCHBERG_1995, NICHOLS_HOLMES_2002}.

We apply TECM to local field potential (LFP) recordings from rat cortex collected before and after a permanent occlusion of the middle cerebral artery \citep{RAT_STROKE_EXPERIMENT_1, RAT_STROKE_EXPERIMENT_2}. TECM identifies which nodes change their role as sources or sinks of information, and which triangular motifs change their recurrent flow, with formal control of false discoveries across the many simultaneous comparisons. In three of four animals, the gradient (hierarchical) component captures a larger share of post-stroke information flow than pre-stroke, indicating a relative shift toward more strictly source-driven propagation at the expense of recurrent feedback; the testing layer flags widespread node- and triangle-level changes accompanying this shift, with the spatial pattern of change varying across subjects. In one animal the affected nodes are spatially organized along the depth axis of the recording array, a pattern suggestive of laminar reorganization that warrants follow-up at the group level. The fourth animal shows no significant changes at either topological scale. While we focus on a neural application, the framework is general: it applies wherever one wishes to characterize directed dependence in a network whose connectivity is intrinsically cyclic.

The remainder of the paper is organized as follows. Section~\ref{sec:meth} reviews the necessary background, introduces lead-lag mutual information for multivariate time series, presents the discrete Hodge decomposition, and describes the permutation-based testing layer. Section~\ref{sec:simu} illustrates and validates the framework on simulated examples with known ground-truth structure. Section~\ref{sec:appl} then applies TECM to the rat LFP recordings collected before and after ischemic stroke, revealing stroke-induced changes in both feed-forward pathways and feedback loops, together with a formal identification of the most significantly affected nodes and triangular motifs. Section~\ref{sec:conc} concludes with a discussion of implications and directions for future work. Per-rat figures and additional methodological details are provided in the Appendix.

\section{Methodology}
\label{sec:meth}

This section introduces the mathematical foundations of TECM. We first define lead-lag mutual information, an information-theoretic quantity that assigns each directed edge a weight quantifying how well the past activity of one channel predicts the present activity of another. We then describe the discrete Hodge decomposition that splits the resulting net information flow into orthogonal components: a gradient flow that is acyclic and thus supports an ordering, and complementary curl and harmonic flows that capture triangle-level and longer-range feedback loops. We close the section with the permutation-based testing procedure used to identify network features that differ significantly between experimental conditions.

\subsection{Information-Theoretic Coupling}
\label{ssec:coupling}

Following standard practice in causal discovery for time series \citep{CI_ASSUMPTIONS_1, CI_ASSUMPTIONS_2}, the construction of a directed dependence network typically proceeds in two stages:
\begin{enumerate}[label=\textit{\roman*)}, leftmargin=*]
    \item \textbf{Structure identification.} Determine the directed skeleton, i.e., the set of ordered pairs $(p,q)$ for which a direct influence $p \to q$ is admissible.
    \item \textbf{Coupling quantification.} Given this skeleton, assign a data-driven weight to each permitted directed edge.
\end{enumerate}
We treat the first stage as a prior informed by domain knowledge and focus on the second. The choice of skeleton is application-dependent: for local electrode arrays one may restrict to spatial neighbors; for whole-brain analyses one may use an anatomical connectivity prior derived from diffusion imaging or include all pairs and rely on the testing layer to suppress spurious edges. For the LFP application of Section~\ref{sec:appl} we restrict candidate edges to spatially adjacent electrodes, mirroring the short-range propagation of cortical activity. The question we address is how strongly channel $p$ predicts the activity of a neighboring channel $q$.

\noindent Any numerical edge weight $w_{pq}$ we assign should satisfy three key requirements:
\begin{enumerate}[label=\textbf{(\alph*)}, leftmargin=*]
    \item \textbf{Directionality.} Discriminate $p \to q$ from $q \to p$.
    \item \textbf{Non-linearity.} Remain sensitive to general, possibly non-linear, interactions.
    \item \textbf{Zero baseline.} Vanish whenever the past of channel $p$ adds no predictive information about channel $q$.
\end{enumerate}
We use mutual information (MI) as our measure of choice. Bivariate MI is a non-negative scalar measure of statistical dependence between two random variables, equal to zero if and only if they are independent \citep{COVER_THOMAS_2006}. We formally define entropy, joint entropy, and MI in Definitions~\ref{def:shannon}--\ref{def:mi} below; here we use $I(A;B)$ as shorthand for the MI between two random variables $A$ and $B$. Bivariate MI is symmetric on its own, but the lead-lag formulation introduced below, which evaluates the past of $p$ against the present of $q$, naturally encodes directionality through temporal asymmetry. It satisfies (b) by construction, and approaches zero at baseline in the sense that $I(X_p(t-k); X_q(t)) = 0$ whenever the two random variables are statistically independent at that lag.

\begin{definition}[Shannon Entropy]
\label{def:shannon}
Let $\mathbf Z$ be a discrete random vector with probability-mass function $p_{\mathbf Z}$. Its entropy is
\[
H(\mathbf Z) \;=\; -\sum_{\mathbf z} p_{\mathbf Z}(\mathbf z) \log p_{\mathbf Z}(\mathbf z),
\]
quantifying the intrinsic uncertainty (or unpredictability) of $\mathbf Z$ in nats.
\end{definition}

\begin{definition}[Joint entropy]
\label{def:joint_ent}
For two discrete random variables $(A,B)$ with joint mass function $p_{A,B}$, the joint entropy is
\[
H(A,B) \;=\; -\sum_{a}\sum_{b} p_{A,B}(a,b) \log p_{A,B}(a,b).
\]
\end{definition}

\begin{definition}[Mutual information]
\label{def:mi}
The mutual information between $A$ and $B$ is
\[
I(A;B) \;=\; H(A) + H(B) - H(A,B),
\]
the average reduction in the uncertainty of $A$ achieved by observing $B$ (and symmetrically for $B$ after observing $A$). It vanishes if and only if $A$ and $B$ are statistically independent.
\end{definition}

\begin{figure}[H]
    \centering
    \includegraphics[width=0.4\linewidth]{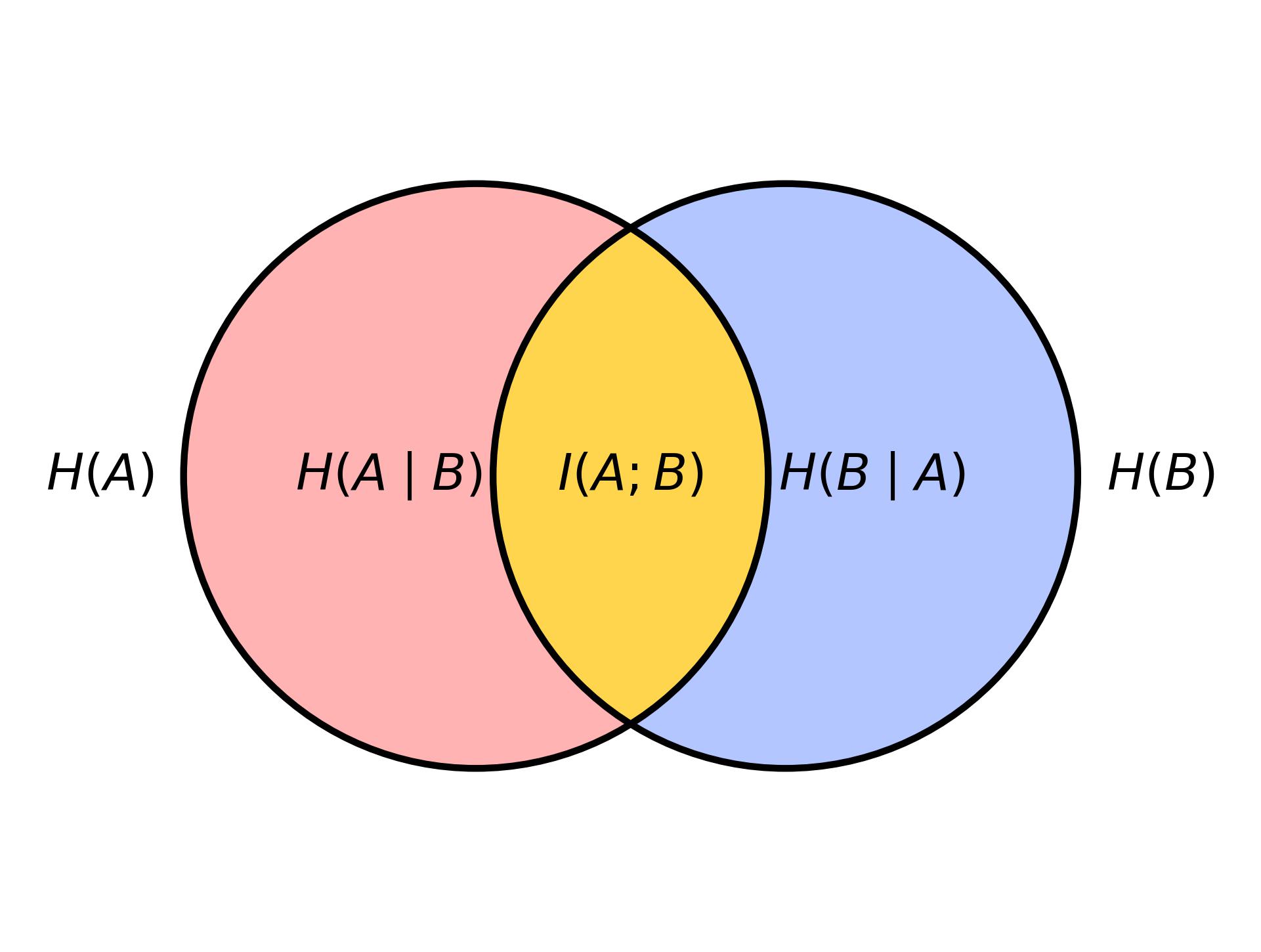}
    \caption{Venn diagram of information content. Circles represent the entropy of each variable; their overlap is the mutual information $I(A;B)$ and the non-overlapping sectors are the conditional entropies.}
    \label{fig:venn_dgm}
\end{figure}

A useful way to visualize these quantities is the Venn diagram in Figure~\ref{fig:venn_dgm}: each circle represents the total entropy of a variable, the non-overlapping sectors are the conditional entropies $H(A\mid B)$ and $H(B\mid A)$, and the overlap is the mutual information $I(A;B)$.

\begin{definition}[Lag-$k$ mutual information]
\label{def:lagged_mi}
Let $\mathbf X(t) = [X_1(t), \dots, X_P(t)]^{\top}$ be a $P$-variate, wide-sense stationary time series observed for $t = 1, \dots, T$. For two distinct channels $p \neq q$ and a positive lag $k$, the lag-$k$ mutual information from $p$ to $q$ is
\[
I_{p\to q}^{(k)} \;=\; I\!\bigl(X_p(t - k)\,;\, X_q(t)\bigr).
\]
\end{definition}

\noindent The asymmetry between $I_{p\to q}^{(k)}$ and $I_{q\to p}^{(k)}$ encodes the directional lead-lag relationship: if the past of $p$ is more informative about the present of $q$ than vice versa, the net flow $W_{p\to q} - W_{q\to p}$ is positive, indicating that $p$ leads $q$.

Based on the lag-specific MI $\{I_{p\to q}^{(k)}\}_{k=1}^{K}$, we define the edge-flow from node $p$ to node $q$ as the cumulative sum
\[
W_{p\to q} \;=\; \sum_{k=1}^K I_{p\to q}^{(k)},
\]
where $K$ is the maximum lag considered. Mathematically, $W_{p\to q}$ is a non-negative scalar obtained by summing $K$ lag-specific MI estimates; because mutual information is non-negative, $W_{p\to q}$ is itself non-negative and is symmetric in $p,q$ only at the level of each individual lag (the directional asymmetry enters through the temporal offset $t-k$ vs. $t$). Biologically, this weight aggregates the information that the past of channel $p$ carries about the present of channel $q$ across delays of physiological interest, producing a single scalar measure of the total influence of $p$ on $q$. We look at the antisymmetric part $W_{p\to q} - W_{q\to p}$, which is the alternating edge flow on which the Hodge decomposition operates (Section~\ref{ssec:hodge}). Because mutual information is not restricted to second-order statistics, $W_{p\to q}$ also captures nonlinear interactions between channels. We estimate $I_{p\to q}^{(k)}$ in practice using the $k$-nearest-neighbor estimator of \citet{KRASKOV_MI}, which is data-efficient and consistent for nonparametric MI estimation.

\subsection{Hodge Decomposition}
\label{ssec:hodge}

Discrete Hodge theory provides a combinatorial analogue of the classical Hodge--Helmholtz decomposition. Over the past decade it has also become a versatile data-analytic tool, particularly for network flows: \citet{HODGE_RANK} introduced HodgeRank, a Hodge-theoretic framework for robust statistical ranking from noisy pairwise comparisons; \citet{HODGE_DNA} used the decomposition to quantify the folding compactness of topologically associating domains in Hi-C chromosome data; and \citet{TDA_ORIENTED_4} applied it to electroencephalography to reveal how epileptic seizures affect different brain regions.

\begin{figure}[H]
    \centering
    \includegraphics[width=\linewidth]{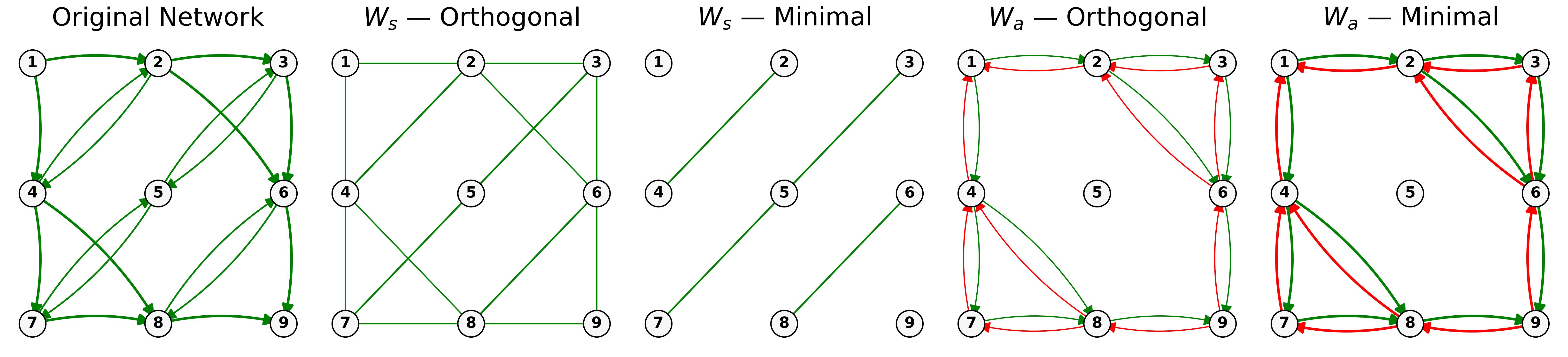}
    \caption{From left to right: original flow $W$, symmetric component $W_s$ from orthogonal projection, symmetric component $W_s$ from minimal projection, skew-symmetric component $W_a$ (orthogonal), and skew-symmetric component $W_a$ (minimal). Green (red) edges denote positive (negative) weights.}
    \label{fig:graph_sym_decomposition}
\end{figure}

The discrete Hodge decomposition requires an alternating edge flow, i.e., weights that obey $w_{pq} = -w_{qp}$ \citep{HODGE_GRAPHS}. An arbitrary weighted digraph $G = (N, E, W)$ must therefore be split into a symmetric part $G_{\mathrm s} = (N, E, W_{\mathrm s})$ and an alternating part $G_{\mathrm a} = (N, E, W_{\mathrm a})$ before the decomposition can be applied (Fig.~\ref{fig:graph_sym_decomposition}). The standard choice for this splitting is the orthogonal projection onto symmetric and skew-symmetric matrices. Below we recall this standard construction, identify an ambiguity it introduces on one-directional edges, and resolve it by adopting a sparsity-oriented alternative that we use throughout the rest of this paper.

\textbf{Orthogonal projection (standard choice).}
Projecting the weight matrix onto the subspaces of symmetric and skew-symmetric matrices yields
\[
W_{s,pq} = \tfrac12\bigl(W_{pq} + W_{qp}\bigr), \qquad W_{a,pq} = \tfrac12\bigl(W_{pq} - W_{qp}\bigr).
\]
This construction minimizes the Frobenius error of each component, ensures orthogonality $\langle W_s, W_a \rangle_F = 0$, and allows exact reconstruction $W = W_s + W_a$. The orthogonal projection does, however, introduce an ambiguity when an edge is purely one-directional: a one-way edge $W_{pq} > 0$, $W_{qp} = 0$ produces a non-zero symmetric entry $W_{s,pq} = W_{s,qp} = W_{pq}/2$ even though no information flows in the $q \to p$ direction, and the skew entry is correspondingly down-scaled to $W_{a,pq} = W_{pq}/2$ (Fig.~\ref{fig:graph_sym_decomposition}, second and fourth panels). For analyses that focus on the alternating flow $W_a$ this scaling is harmless up to a factor of two, but for studies that wish to interpret the symmetric layer $W_s$ as ``shared bidirectional content'' the ambiguity is consequential. We resolve it as follows.

\textbf{Minimal (sparsity-oriented) decomposition.}
To remove this ambiguity we adopt a minimal split that preserves one-directional edges intact:
\[
W_{s,pq} = \min\bigl\{W_{pq}, W_{qp}\bigr\}, \qquad W_{a,pq} = W_{pq} - W_{qp}.
\]
Here $W_{s,pq}$ is non-zero only when edges exist in both directions; pure one-way links stay entirely in $W_a$ at full strength, instead of being halved. The symmetric layer is therefore much sparser and the alternating layer keeps the original magnitudes (Fig.~\ref{fig:graph_sym_decomposition}, third and fifth panels). The trade-off is that strict additivity is lost: in general $W \neq W_s + W_a$ at the matrix level, though both layers retain a clean interpretation. We adopt this minimal decomposition throughout the rest of the paper. We emphasize that this paper focuses on the alternating component $W_a$, which feeds the Hodge decomposition; the analysis of $W_s$ in its own right is left for future work.

Two complementary viewpoints from TDA are useful here. Persistent homology (PH) tracks how $k$-dimensional ``holes'' appear and disappear across a filtration \citep{TDA_INTRO_YAAGOUBI}; it is formulated on the chain complex with boundary maps. Discrete Hodge theory (HD) works instead on the cochain complex, which is the Hilbert-space dual of the chain complex. Equipped with an inner product, HD decomposes flows into exact (gradient), coexact (curl), and harmonic parts \citep{HODGE_GRAPHS}. In this work we restrict to a 2-D simplicial complex (nodes, edges, and triangular faces), so the cochain complex terminates at degree 2:

\[
\begin{array}{r c c c c c}
\text{chains:}   & C_2 & \xrightarrow{\;\partial_2\;} & C_1 & \xrightarrow{\;\partial_1\;} & C_0 \\[4pt]
\text{cochains:} & C^{2} & \xleftarrow{\;\delta_{1} = \partial_{2}^{\!*}\;} & C^{1} & \xleftarrow{\;\delta_{0} = \partial_{1}^{\!*}\;} & C^{0}.
\end{array}
\]

Here $C_k$ is the real vector space of $k$-chains, i.e., formal linear combinations of $k$-simplices (triangles, edges, nodes for $k = 2, 1, 0$), and $C^{k} = (C_k)^{\!*}$ is its dual cochain space, whose elements assign a real value to each $k$-simplex. Concretely, $C^0$ assigns a scalar potential to each node, $C^1$ assigns a signed flow to each oriented edge, and $C^2$ assigns a circulation value to each oriented triangle. The boundary $\partial_k : C_k \to C_{k-1}$ maps an oriented $k$-simplex to the signed sum of its $(k-1)$-faces; with the chosen inner products, the coboundary is the adjoint $\delta_k = \partial_{k+1}^{\!*} : C^{k} \to C^{k+1}$. For our 2-D complex, $\delta_0 : C^{0} \to C^{1}$ is the node-edge incidence (the discrete \textbf{gradient} on scalar node potentials), while $\delta_1 : C^{1} \to C^{2}$ accumulates an edge flow around each triangle (the discrete \textbf{curl}). In PH one studies homology $H_k = \ker(\partial_k) / \mathrm{im}(\partial_{k+1})$; in HD, the edge cochain space $C^{1}$ (where our alternating flow $W_a$ lives) admits the orthogonal Hodge splitting
\[
C^{1} \;=\; \mathrm{im}(\delta_{0}) \;\oplus\; \mathrm{im}(\delta_{1}^{\!*}) \;\oplus\; \bigl(\ker \delta_{1} \cap \ker \delta_{0}^{\!*}\bigr),
\]
corresponding to gradient, curl, and harmonic components \citep{HODGE_RANK}. Accordingly, any alternating edge flow in $C^{1}$ (i.e., $W_a \in \mathbb{R}^{|E|}$) splits uniquely as
\[
W_a \;=\; -\delta_{0}\,\varphi \;+\; \delta_{1}^{\!*}\,\psi \;+\; h,
\]
with a node potential $\varphi \in C^0$ (interpretable as hierarchical ranks of nodes), a rotational flow $\psi \in C^2$ (curl term), and a harmonic component $h \in \ker(\delta_{1}) \cap \ker(\delta_{0}^{\!*})$; that is, $h$ is curl-free and source-free, yet supports nontrivial circulation around topological holes.

\subsection{Estimating Hodge Components}
\label{ssec:estimate}

Given an alternating edge flow $W_a \in C^{1}$ and the decomposition above, we estimate the potentials $\varphi \in C^{0}$ (weights on the nodes) and $\psi \in C^{2}$ (weights on the triangles) by least-squares projections onto $\mathrm{im}(\delta_{0})$ and $\mathrm{im}(\delta_{1}^{\!*})$, respectively. With the standard $\ell^2$ inner product on $C^{1}$, this amounts to the convex quadratic problems
\[
\widehat{\varphi} \;\in\; \arg\min_{\varphi \in C^{0}} \;\Big\| W_a + \delta_{0}\varphi \Big\|_{F}^{2}, \qquad
\widehat{\psi} \;\in\; \arg\min_{\psi \in C^{2}} \;\Big\| W_a - \delta_{1}^{\!*}\psi \Big\|_{F}^{2}.
\]
For an oriented edge $p \to q$,
\[
(\delta_{0}\varphi)_{pq} \;=\; \varphi(q) - \varphi(p), \qquad
(\delta_{1}^{\!*}\psi)_{pq} \;=\; \sum_{\tau \in T : (p,q) \subset \tau} \mathrm{sgn}\bigl((p,q), \tau\bigr) \, \psi_{\tau},
\]
i.e.\ $\delta_{1}^{\!*}$ sums the triangle potentials incident to $(p,q)$ with the appropriate orientation signs. Since $C^{1}$ splits as an orthogonal direct sum, the harmonic component is the orthogonal residual after projecting onto the gradient and curl subspaces:
\[
\widehat{h} \;=\; W_a + \delta_{0}\widehat{\varphi} - \delta_{1}^{\!*}\widehat{\psi},
\]
which by construction lies in $\ker \delta_{1} \cap \ker \delta_{0}^{\!*}$.

In practice the two least-squares problems are convex quadratics with either low dimensionality (vector unknowns) or sparse structure, and can be solved efficiently. For small to moderate networks we use a generic convex optimization solver. For higher-dimensional or larger problems we can exploit the problem's structure by setting the gradient of the objective to zero, yielding the normal equations
\[
\Delta_{0}\,\widehat{\varphi} \;=\; -\delta_{0}^{\!*} W_a, \qquad \Delta_{2}\,\widehat{\psi} \;=\; \delta_{1} W_a,
\]
where
\[
\Delta_{0} \;:=\; \delta_{0}^{\!*}\delta_{0} \quad \text{(node Laplacian)}, \qquad \Delta_{2} \;:=\; \delta_{1}\delta_{1}^{\!*} \quad \text{(face Laplacian)}.
\]
Both systems are symmetric and positive semidefinite, with null spaces corresponding to constant potentials on each connected component (gauge freedom). We fix the gauge either by imposing a mean-zero constraint on the potentials or by adding a small ridge penalization (Tikhonov regularization), which also acts as a preconditioner for iterative solvers.

\subsection{Hypothesis Testing for Between-Condition Changes}
\label{ssec:testing}

A key goal of our analysis is not merely to describe the information-flow network but to identify which of its structural features change between conditions, in our application the pre- and post-stroke epochs, in a statistically rigorous way. Testing the raw edge flow $W_a$ edge by edge would keep us at the level of pairwise interactions: each test asks only whether the link between two specific nodes has changed. This is precisely the pairwise view that motivated the higher-order perspective of this paper. The Hodge decomposition gives us quantities that are higher-order by construction, and these are what we test. The node potential $\varphi_p \in C^0$ is not a property of any single edge but of node $p$'s position in the feed-forward hierarchy relative to the entire network; it summarizes whether $p$ acts overall as a source or a sink of information, so a change in $\varphi_p$ between conditions means $p$'s global sending-versus-receiving role has shifted. The curl potential $\psi_\tau \in C^2$ is a genuinely three-node quantity: it measures the recurrent circulation around triangle $\tau$, including the sense in which information cycles among its three nodes, so a change in $\psi_\tau$ means that local loop has been reorganized in magnitude or reversed in direction. Testing $\varphi$ and $\psi$ therefore asks directly which hierarchical roles and which recurrent loops the stroke alters, higher-order questions that no pairwise edge test can pose. We therefore develop a permutation-based testing procedure \citep{NICHOLS_HOLMES_2002} that operates at two topological levels: the node level, with test quantity $\varphi_p$, and the triangle level, with test quantity $\psi_\tau$.

\textbf{Test statistics.}
Let $\{\widehat{\varphi}_p^{(m)}\}_{m=1}^{M_{\mathrm{pre}}}$ and $\{\widehat{\varphi}_p^{(m)}\}_{m=M_{\mathrm{pre}}+1}^{M_{\mathrm{pre}} + M_{\mathrm{post}}}$ denote the estimated node potentials for node $p$ across all pre- and post-stroke epochs. We summarize the change at node $p$ by the standardized mean difference
\[
T_p^{\varphi} \;=\; \frac{\bar\varphi_p^{\mathrm{post}} - \bar\varphi_p^{\mathrm{pre}}}{s_p^{\varphi}},
\]
where $\bar\varphi_p^{\mathrm{pre/post}}$ are epoch-averaged potentials and $s_p^{\varphi}$ is the pooled standard error across epochs. An analogous statistic $T_\tau^{\psi}$ is defined for each triangle $\tau$ using its curl potentials $\{\widehat{\psi}_\tau^{(m)}\}$.

\textbf{Permutation null distribution.}
Under the null hypothesis that the condition (pre vs.\ post) has no effect on node $p$ (respectively, triangle $\tau$), the epoch labels are exchangeable. We obtain a null distribution $\{T_p^{\varphi,(b)}\}_{b=1}^B$ by reassigning epoch labels and recomputing the relevant test statistic for each reassignment. When the number of distinct label assignments is small, we enumerate all of them exactly; otherwise we sample $B$ of them at random. The two-sided permutation $p$-value for node $p$ is
\[
\mathrm{pval}_p \;=\; \frac{1 + \sum_{b=1}^{B} \mathbf{1}\!\bigl(|T_p^{\varphi,(b)}| \geq |T_p^{\varphi}|\bigr)}{1 + B},
\]
with the analogous definition for triangles. Because successive epochs are not fully independent, we adopt a block permutation scheme that permutes contiguous blocks of epochs rather than individual epochs, thereby preserving short-range temporal autocorrelation \citep{KUNSCH_1989, POLITIS_ROMANO_1994}. The block length and number of permutations are reported per experiment in the relevant sections; in the application we use a block length of $6$ epochs and $B = 5{,}000$ permutations.

\textbf{Multiple comparisons.}
With $|N|$ nodes and $|T|$ triangles tested simultaneously, controlling the familywise error rate (FWER) via Bonferroni correction is straightforward but conservative given the strong spatial dependence among electrodes. We therefore primarily report results using the Benjamini--Hochberg (BH) procedure to control the false discovery rate (FDR) at level $\alpha = 0.05$ \citep{BENJAMINI_HOCHBERG_1995}; this procedure retains validity under positive regression dependence \citep{BENJAMINI_YEKUTIELI_2001}, which is the relevant dependence structure on a regular grid.

\section{Simulations}
\label{sec:simu}

\begin{figure}[H]
    \centering
    \includegraphics[width=0.85\linewidth]{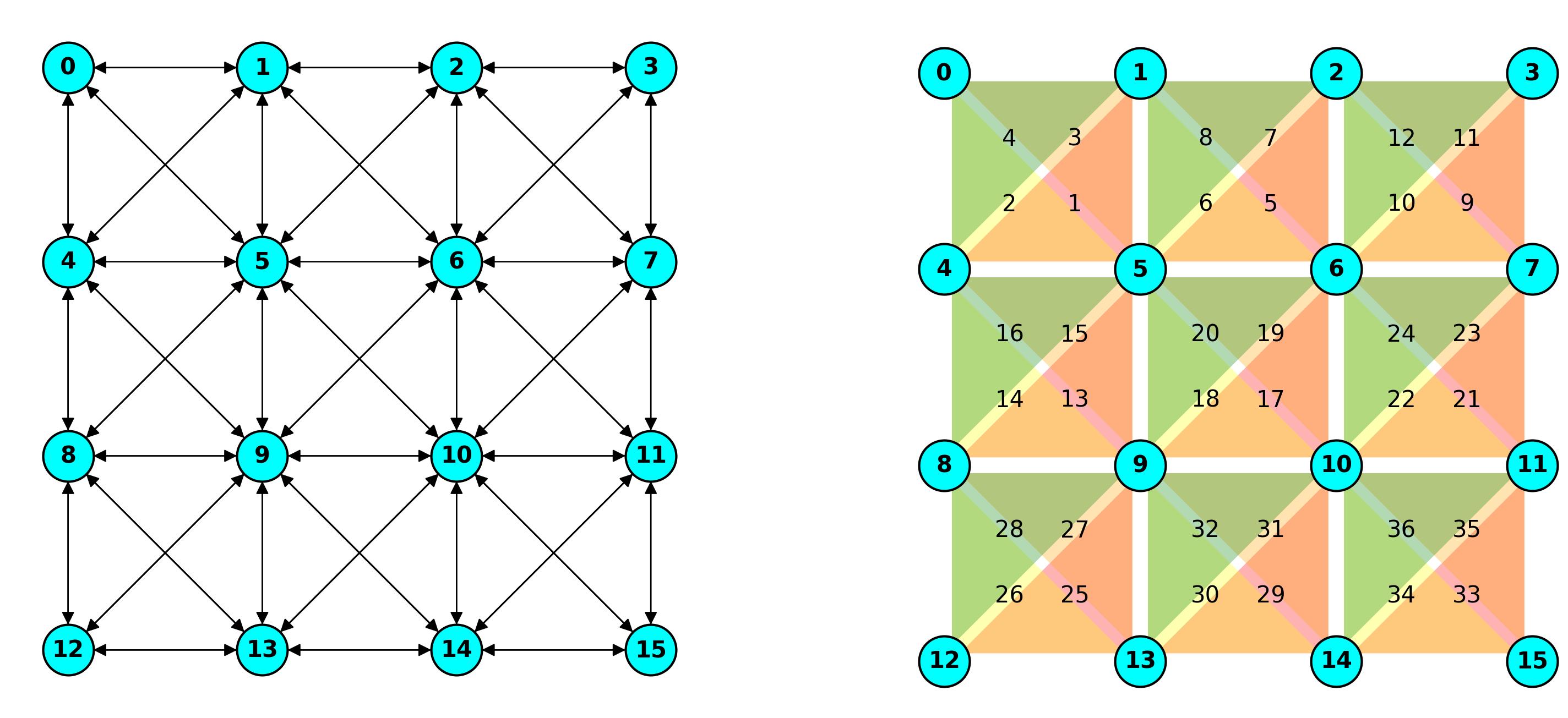}
    \caption{The $4\times4$ toy grid used for validation. Left: directed graph with bidirectional edges ($42$ undirected edges). Right: triangulated 2-simplicial complex ($36$ oriented triangles).}
    \label{fig:simu_toy_grid}
\end{figure}

Before turning to real recordings, we test TECM on synthetic data where the ground truth is known. All experiments use a small network of $16$ channels arranged on a $4\times4$ grid, small enough to inspect every node and triangle by eye yet rich enough to support all three Hodge components. We run four experiments in two groups. The first two check the decomposition itself: we build edge flows by hand from a known gradient and curl and confirm they are recovered exactly, then add a topological hole and a harmonic circulation and confirm all three components are recovered. The next two check the inference layer: we plant a change in one node's role and confirm the test detects it, then generate data with no change between conditions and confirm the test stays silent. The grid carries horizontal, vertical, and diagonal bidirectional edges ($42$ undirected edges in total) and a full triangulation of each unit square ($36$ oriented triangles), shown in Figure~\ref{fig:simu_toy_grid}.

\subsection{Ground-Truth Recovery}
\label{ssec:simu_recovery}

We begin on the fully triangulated grid. Here every cycle bounds a combination of triangular faces, so the first homology is trivial and the harmonic subspace is zero-dimensional: any alternating flow is exactly gradient plus curl, with no harmonic part. The harmonic term is not absent from the framework, only from this particular complex; we exhibit it on a holed complex in the next subsection.

\begin{figure}[H]
    \centering
    \includegraphics[width=0.9\linewidth]{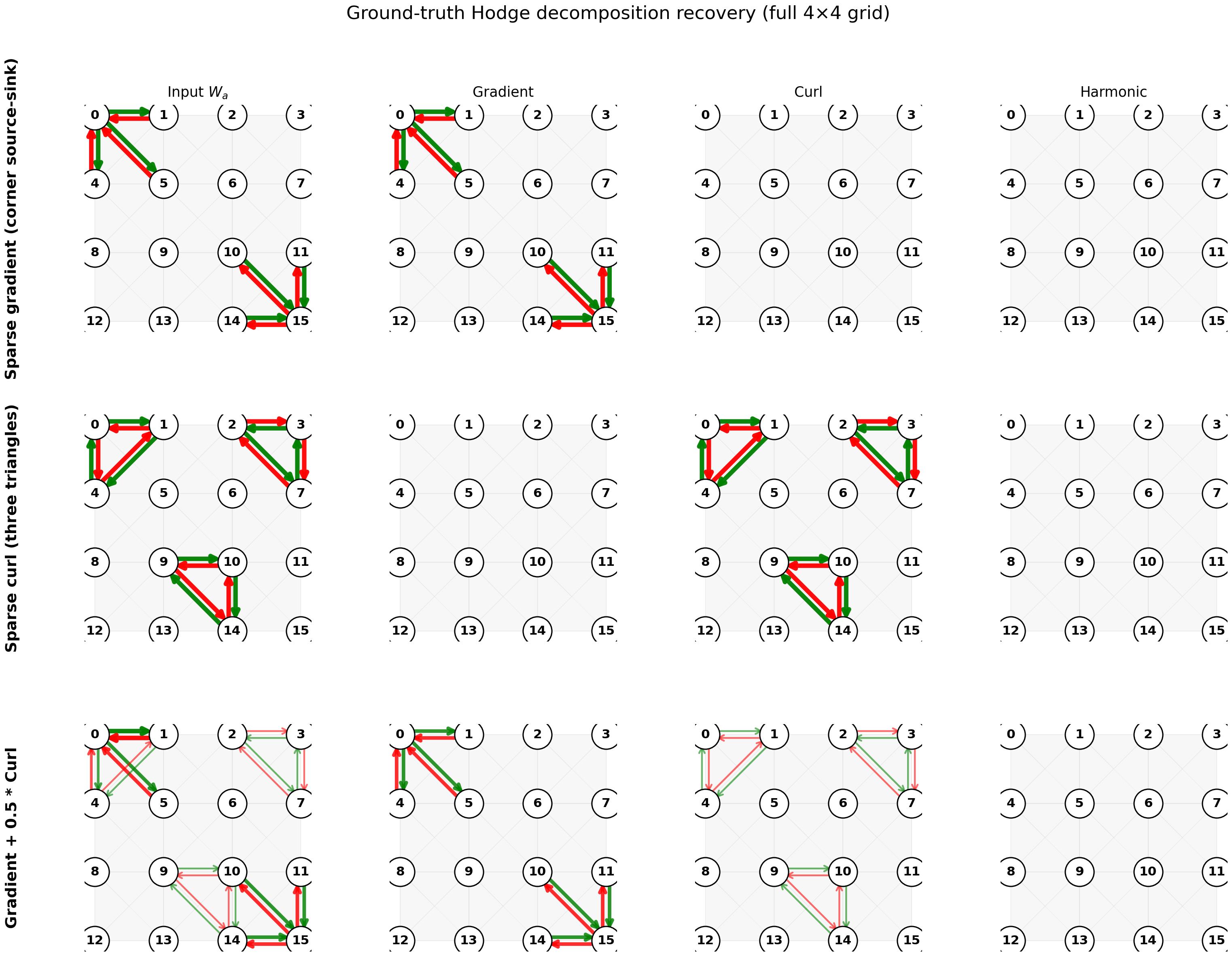}
    \caption{Hodge recovery on the fully triangulated grid. The harmonic column is empty by construction, since the harmonic subspace here is zero-dimensional. Rows: pure gradient, pure curl, mixed input. Columns: input flow $W_a$, recovered gradient $-\delta_0 \widehat\varphi$, curl $\delta_1^{\!*}\widehat\psi$, and harmonic $\widehat h$. Arrow color encodes sign (green positive, red negative) and width encodes magnitude.}
    \label{fig:simu_recovery}
\end{figure}

To check recovery, we build three flows with known structure. Let $\varphi^{\star}$ be a sparse node potential placed on two opposite corners ($+1.5$ at node $0$, $-1.5$ at node $15$) and centered to zero mean, and let $\psi^{\star}$ be a sparse triangle potential supported on three triangles in distinct regions with mixed signs ($+1$, $-1$, $+1$). From these we form a pure gradient $W_{\text{grad}}$, a pure curl $W_{\text{curl}}$, and a mixture $W_{\text{mix}} = W_{\text{grad}} + 0.5\,W_{\text{curl}}$, then decompose each. Table~\ref{tab:simu_recovery} reports the resulting component norms and Figure~\ref{fig:simu_recovery} shows the edge-wise reconstructions.

\begin{table}[H]
\centering
\caption{Ground-truth recovery on the fully triangulated $4\times4$ grid. The harmonic norm sits at machine precision in every case because this grid has trivial first homology; the next subsection exhibits a genuine harmonic component on a complex with a hole.}
\label{tab:simu_recovery}
\begin{tabular}{lcccc}
\toprule
Input & $\|W_a\|$ & $\|\text{grad}\|$ & $\|\text{curl}\|$ & $\|\text{harm}\|$ \\
\midrule
Pure gradient (sparse corner source-sink)  & $3.674$ & $3.674$ & $0.000$ & $3.1 \times 10^{-8}$ \\
Pure curl (three triangles)                & $3.000$ & $0.000$ & $3.000$ & $7.8 \times 10^{-8}$ \\
Gradient $+\,0.5\,\times\,$curl            & $3.969$ & $3.674$ & $1.500$ & $5.0 \times 10^{-8}$ \\
\bottomrule
\end{tabular}
\end{table}

The recovery is exact to machine precision. The pure-gradient input lands entirely in the gradient component, the pure-curl input entirely in the curl component, and the mixture splits in the correct $1 : 0.5$ ratio of gradient to curl. The harmonic norm is zero (to numerical tolerance) throughout, as it must be on a complex with no holes, and the edge-wise maps in Figure~\ref{fig:simu_recovery} confirm that the off-target components vanish.

\subsection{The Harmonic Component on a Complex with a Hole}
\label{ssec:simu_harmonic}

\begin{figure}[H]
    \centering
    \includegraphics[width=0.9\linewidth]{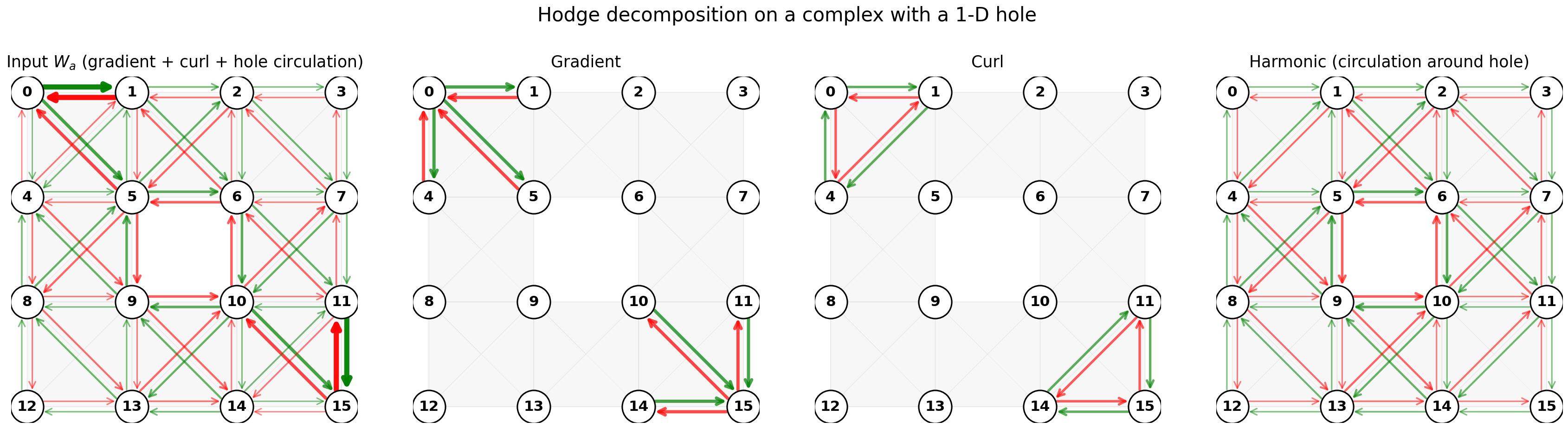}
    \caption{Hodge decomposition on a $4\times4$ grid with the central square left un-triangulated, creating a 1-D hole. Left to right: mixed input flow (gradient $+$ curl $+$ hole circulation); recovered gradient; recovered curl; recovered harmonic, circulating around the hole.}
    \label{fig:simu_harmonic}
\end{figure}

To bring the harmonic component to life we put a hole in the grid. Removing the four triangles around the central unit square, while keeping their bounding edges, leaves a loop of four edges that encloses no face. This is a genuine 1-D hole, and it makes the harmonic subspace one-dimensional. We then build an input flow with all three ingredients present: a sparse gradient (a source-sink pair on nodes $0$ and $15$), a sparse curl (two distant triangles rotating in opposite senses), and a pure harmonic mode, obtained as the null-space eigenvector of the edge Laplacian $L_1 = B_1 B_1^{\top} + B_2^{\top} B_2$ and rescaled to magnitude $3.5$. Decomposing this flow gives
\[
\|W_a\| = 5.635, \qquad \|\text{grad}\| = 3.674, \qquad \|\text{curl}\| = 2.449, \qquad \|\text{harm}\| = 3.500,
\]
each matching its planted value, and the three norms satisfy the Pythagorean identity $\sqrt{\|\text{grad}\|^2 + \|\text{curl}\|^2 + \|\text{harm}\|^2} = 5.635 = \|W_a\|$ expected from the orthogonality of the subspaces. Figure~\ref{fig:simu_harmonic} shows the split: the harmonic component traces a clean circulation around the hole, visibly distinct from the local triangle-level circulation of the curl term and the source-sink pattern of the gradient. This is exactly the structure the harmonic term exists to capture. It plays no role on the fully triangulated LFP grid of Section~\ref{sec:appl}, whose first homology is again trivial, but it would become essential for any analysis on a complex with holes.

\subsection{Detection of a Known Hierarchical Perturbation}
\label{ssec:simu_detection}

We now test the inference layer of Section~\ref{ssec:testing} on data with a planted change. On the same grid we simulate a first-order vector autoregressive process $\mathbf X(t) = A\,\mathbf X(t-1) + \boldsymbol\varepsilon(t)$ with $\boldsymbol\varepsilon(t) \sim \mathcal N(0, \sigma^2 I)$, where the coupling matrix $A$ has self-decay $0.5$ on the diagonal and a weak baseline coupling of $0.05$ on every directed edge. The two conditions differ in one place only: in the post condition, the four edges feeding node $5$ from its neighbors $\{1, 4, 6, 9\}$ are strengthened from $0.05$ to $0.70$, turning node $5$ into a strong sink. For each condition we draw $M = 30$ independent epochs of $T = 1000$ samples, estimate lead-lag MI networks at $k_{\max} = 3$, compute Hodge potentials, and run the block-permutation test (block length $5$) with BH-FDR at $\alpha = 0.05$. Because $30$ epochs split into six blocks per condition, only $\binom{12}{6} = 924$ distinct block assignments exist, so the test enumerates all of them exactly rather than sampling.

\begin{figure}[H]
    \centering
    \includegraphics[width=0.9\linewidth]{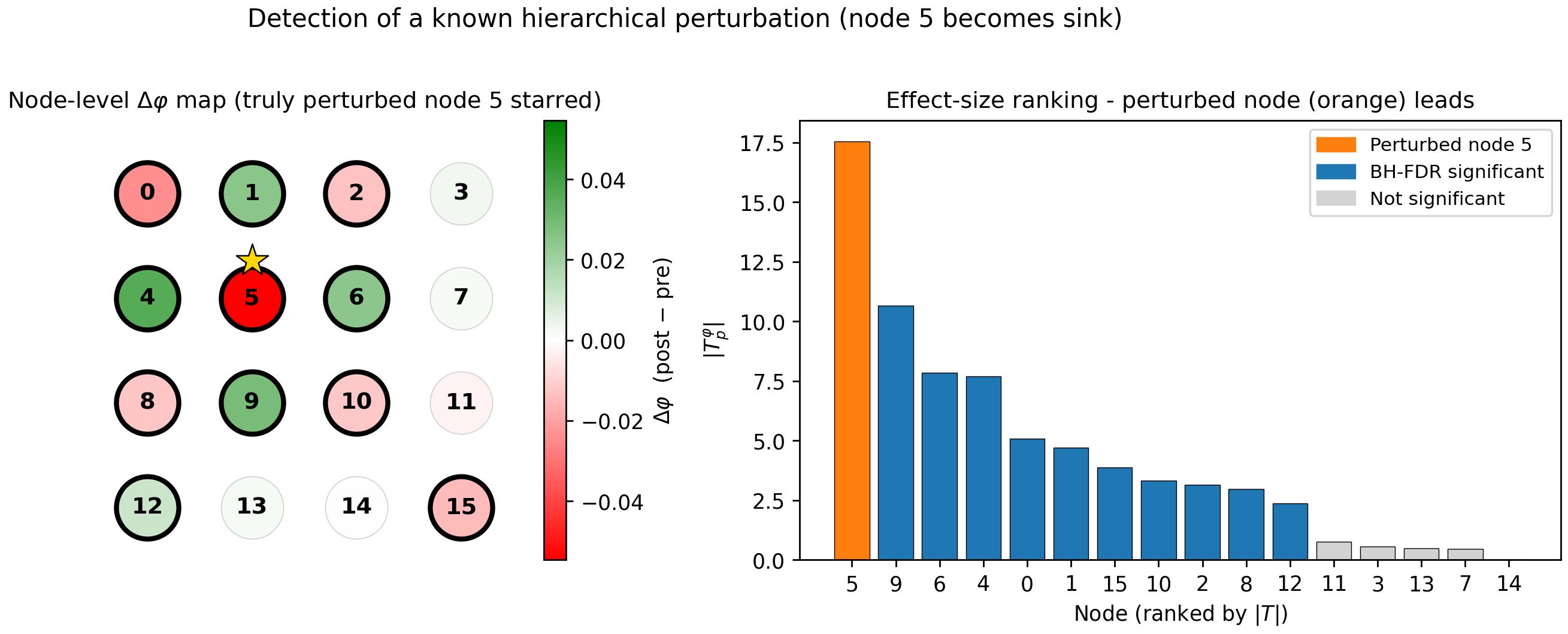}
    \caption{Detection of a planted hierarchical perturbation on the VAR system. Left: spatial map of $\Delta\widehat\varphi$; the perturbed node ($5$, gold star) shows the largest negative shift and its four driving neighbors the largest positive shifts. Right: $|T_p^\varphi|$ ranked by effect size.}
    \label{fig:simu_detection}
\end{figure}

The test recovers the perturbed node squarely at the top. Table~\ref{tab:simu_detection} reports the full ranking by effect size $|T_p^\varphi|$: node $5$ leads at $|T| = 17.55$, well clear of the runner-up ($|T| = 10.66$ at node $9$). The spatial picture in Figure~\ref{fig:simu_detection} shows node $5$ in deep red, ringed by its four driving neighbors $\{1, 4, 6, 9\}$ in green, the signature of a freshly created sink.

\begin{table}[H]
\centering
\caption{Node-level results for the detection experiment, ranked by $|T_p^\varphi|$ (all $924$ block permutations enumerated exactly). The truly perturbed node is node $5$.}
\label{tab:simu_detection}
\begin{tabular}{rrcrrc}
\toprule
Rank & Node & $\Delta\widehat\varphi$ & $|T|$ & $q_{\text{BH}}$ & Reject \\
\midrule
$1$  & $\mathbf{5}$  & $-0.0546$ & $\mathbf{17.55}$ & $0.0074$ & \checkmark \\
$2$  & $9$           & $+0.0290$ & $10.66$ & $0.0074$ & \checkmark \\
$3$  & $6$           & $+0.0243$ & $7.85$  & $0.0074$ & \checkmark \\
$4$  & $4$           & $+0.0362$ & $7.69$  & $0.0074$ & \checkmark \\
$5$  & $0$           & $-0.0239$ & $5.08$  & $0.0074$ & \checkmark \\
$6$  & $1$           & $+0.0250$ & $4.71$  & $0.0074$ & \checkmark \\
$7$  & $15$          & $-0.0146$ & $3.87$  & $0.0108$ & \checkmark \\
$8$  & $10$          & $-0.0119$ & $3.32$  & $0.0393$ & \checkmark \\
$9$  & $2$           & $-0.0129$ & $3.14$  & $0.0363$ & \checkmark \\
$10$ & $8$           & $-0.0122$ & $2.97$  & $0.0074$ & \checkmark \\
$11$ & $12$          & $+0.0114$ & $2.37$  & $0.0363$ & \checkmark \\
$12$ & $11$          & $-0.0029$ & $0.76$  & $0.6746$ & $-$        \\
$13$ & $3$           & $+0.0031$ & $0.55$  & $0.6427$ & $-$        \\
$14$ & $13$          & $+0.0023$ & $0.49$  & $0.6184$ & $-$        \\
$15$ & $7$           & $+0.0018$ & $0.46$  & $0.6746$ & $-$        \\
$16$ & $14$          & $-0.0000$ & $0.00$  & $0.9957$ & $-$        \\
\bottomrule
\end{tabular}
\end{table}

Beyond the perturbed node and its four neighbors, eleven of the $16$ nodes clear the BH threshold, with significant shifts reaching sites such as $0$, $2$, $8$, $10$, $12$, and $15$ that share no edge with node $5$. This spread is the higher-order signature the framework is designed to expose: a localized change in pairwise coupling reshapes the directed-flow geometry across the whole grid, and the gradient component summarizes that network-wide reorganization as a single $|N|$-dimensional vector of hierarchical positions rather than $|E|$ correlated edge p-values. The ordering by effect size is what matters operationally, and the truly perturbed node sits unambiguously at the top, just as we rely on it to in the application (Section~\ref{ssec:appl_testing}). The triangle-level test, run in parallel, flags $15$ of $36$ triangles, reflecting the same network-wide reorganization of circulation around the new sink.

To confirm that the test stays quiet when no change is present, we run a companion null experiment: $R = 500$ Monte-Carlo trials in which both conditions are generated from the same coupling matrix (self-decay $0.5$, baseline coupling $0.10$), so any rejection is by definition a false positive. Each trial uses $M = 20$ epochs per condition at $T = 500$ samples; with four blocks per condition the test enumerates all $\binom{8}{4} = 70$ block assignments exactly, and BH-FDR at $\alpha = 0.05$ is applied separately to the node and triangle tests. Across all $500$ trials the procedure rejects exactly zero nodes and zero triangles, confirming that the breadth of rejections in the detection experiment above reflects genuine propagation of the perturbation rather than liberal behavior of the test.

The four experiments together show that the decomposition recovers each of its components when the ground truth is known, that the harmonic term captures circulation whenever the complex has holes, that the test ranks a genuine perturbation at the top, and that it produces no false positives under the global null. We now apply the same pipeline to LFP recordings before and after experimental stroke.

\section{Application}
\label{sec:appl}

Ischemic stroke remains one of the leading causes of death and long-term disability worldwide. It occurs when a clot or arterial blockage reduces or interrupts blood flow to a region of the brain, depriving neurons of oxygen and nutrients. The damage that follows can be rapid and irreversible, producing an infarct in the affected cortical or subcortical tissue. The resulting functional impairment depends on the size and location of the infarct and on the brain's capacity to reorganize around it \citep{ISCHEMIC_STROKE_RESEARCH}.

Neuroimaging is essential for diagnosing stroke and assessing its anatomical extent \citep{ISCHEMIC_STROKE_IMAGING_1, ISCHEMIC_STROKE_IMAGING_2}. Imaging modalities give a macroscopic view of the damaged territory and inform prognosis, but they cannot directly resolve the fine-scale electrophysiological changes that take place in the affected networks. Experimental rodent models complement imaging by offering precisely the temporal and spatial resolution that imaging lacks. We analyze LFP recordings from a rodent model of acute ischemic stroke, in which a permanent occlusion of the distal middle cerebral artery (pMCAo) produces a focal cortical infarct in the posterior medial barrel subfield (PMBSF), a subregion of primary somatosensory cortex (S1) within MCA territory \citep{RAT_STROKE_EXPERIMENT_1, RAT_STROKE_EXPERIMENT_2}. The stroke induction acts as an experimental perturbation at the level of the brain itself, and our analysis below compares the directed dependence networks estimated before and after this perturbation to identify which features change at multiple topological scales.

\begin{figure}[H]
    \centering
    \includegraphics[width=\linewidth]{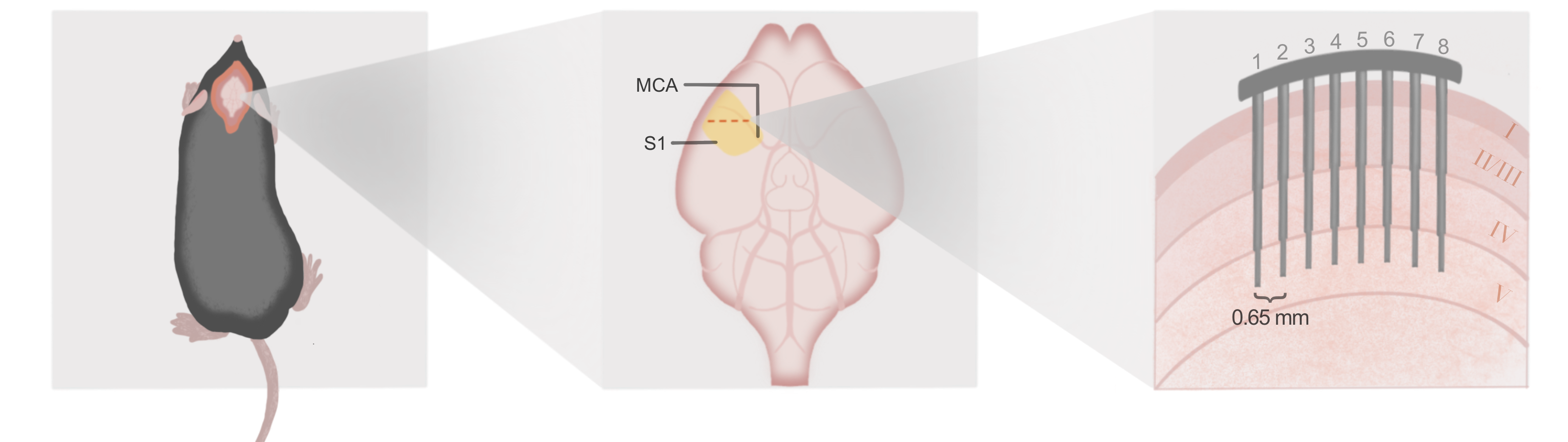}
    \caption{Experimental setup for the ischemic stroke model (pMCAo) with LFP recordings in S1 near the MCA territory. Left: schematic of the rat during pMCAo. Middle: dorsal view of MCA territory and S1. Right: cortical slice showing electrode placement across layers.}
    \label{fig:stroke_lfp_setting}
\end{figure}

\begin{figure}[H]
    \centering
    \includegraphics[width=0.9\linewidth]{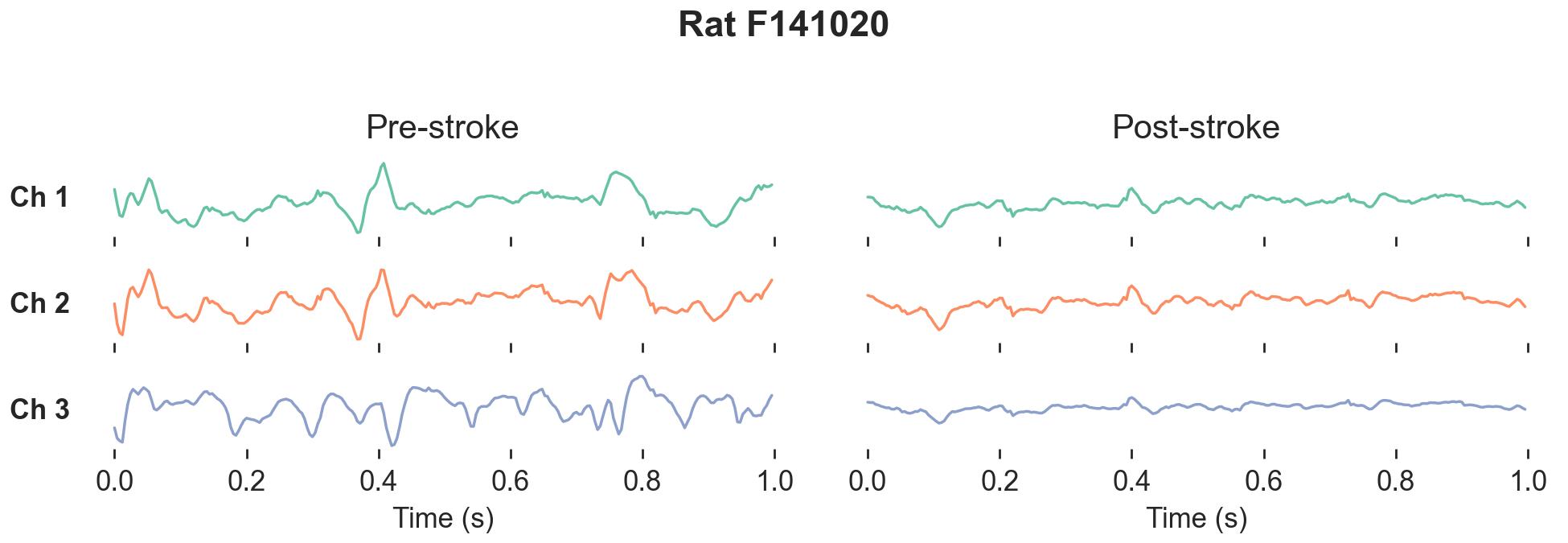}
    \caption{Sample LFP recordings from channels 1--3 in one rat. Each trace shows a representative 1\,s epoch before and after stroke. Data acquired at 1\,kHz and downsampled to 250\,Hz.}
    \label{fig:lfp_data}
\end{figure}

A 32-channel microelectrode array was implanted across cortical layers within the MCA territory, spanning S1 and adjacent regions. The array comprises eight horizontally aligned recording sites spaced $0.65\,\mathrm{mm}$ apart and four recording depths (300, 700, 1100, and $1500\,\mu\mathrm{m}$), sampling cortical layers II/III through Vb (Fig.~\ref{fig:stroke_lfp_setting}). After implantation, electrodes settled for approximately one hour before baseline acquisition. Continuous LFP recordings were then obtained before and after pMCAo; representative pre- and post-occlusion traces are shown in Fig.~\ref{fig:lfp_data}. Importantly, the four rows of the array correspond to cortical depth rather than horizontal position, so any vertical structure we observe in the network analysis below admits a direct laminar interpretation.

Previous analyses of this dataset showed that, within minutes of pMCAo, spontaneous LFPs exhibit a rapid and widespread increase in spatiotemporal synchrony across cortical depths and horizontal locations, driven by bursts of low-frequency, multi-band oscillations. This heightened synchrony persists throughout the acute ischemic period while evoked responses remain largely preserved \citep{RAT_STROKE_EXPERIMENT_1}. Sham controls show no comparable changes. Follow-up experiments demonstrated that early sensory stimulation delivered immediately after occlusion attenuates this synchrony increase, while delayed stimulation does not, identifying the synchrony increase as a candidate early biomarker of ischemic progression \citep{RAT_STROKE_EXPERIMENT_2}.

\begin{figure}[H]
    \centering
    \includegraphics[width=0.8\linewidth]{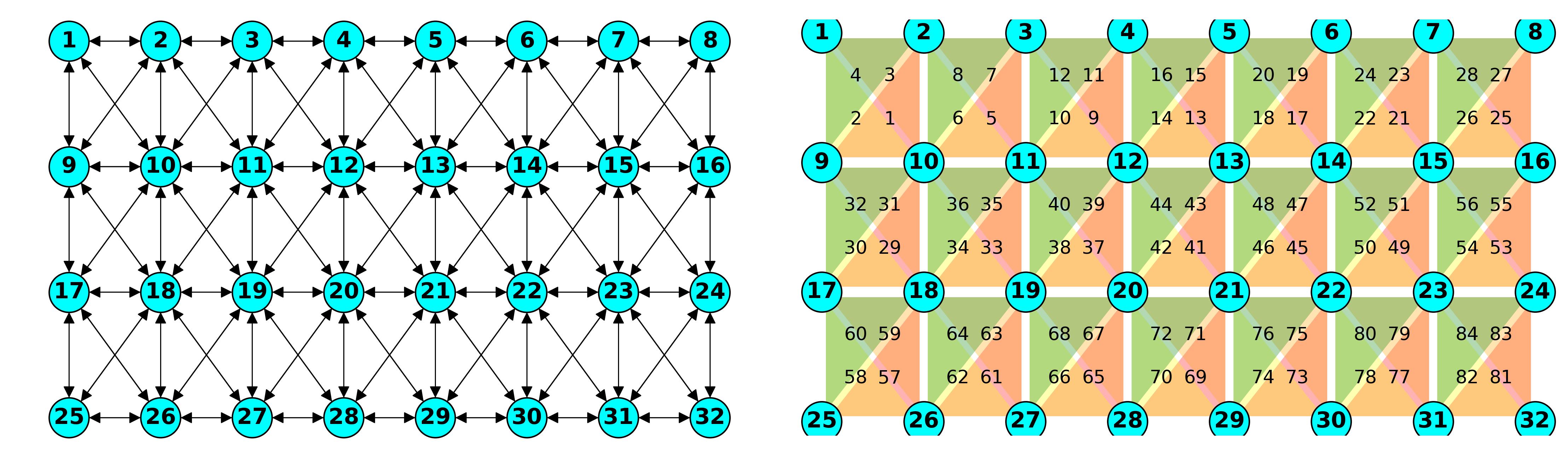}
    \caption{The 2-D simplicial complex for the LFP grid. Left: directed network with edges between neighboring electrodes. Right: triangular faces tiling local neighborhoods.}
    \label{fig:lfp_grid_complex}
\end{figure}

These earlier analyses relied on correlation or coherence, which yield symmetric connectivity matrices and therefore conflate direction-of-flow with mere co-fluctuation. Such measures capture overall synchrony well but cannot tell us which site is driving and which is being driven, nor whether an apparent pairwise change reflects a direct interaction or one mediated by a third site. The TECM framework directly addresses both of these limitations. We estimate lead-lag mutual information between electrode pairs to quantify directed statistical dependence, with temporal shifting providing the asymmetry required to recover direction. The resulting weighted, directed network represents putative directed dependence among recording sites. We treat the LFP array as a 2-D simplicial complex (Fig.~\ref{fig:lfp_grid_complex}): vertices (electrodes), edges linking spatially adjacent sites, and oriented triangles tiling local neighborhoods of the recording grid. Restricting edges to adjacent sites reflects the geometry of this recording array, which spans a confined cortical region; long-range interactions with regions outside the array (e.g., interhemispheric or thalamocortical) are not measured in this experiment and would require either a wider array or an anatomically informed prior on inter-regional connectivity. Because triangles are spatially restricted by construction (Fig.~\ref{fig:lfp_grid_complex}), the curl component reported below captures feedback among physically neighboring electrodes only. The lead-lag weights provide a directed edge flow on this complex; the discrete Hodge decomposition then splits this edge flow into a gradient (feed-forward) component and a curl (triangle-level feedback) component, and yields node-level potentials $\widehat{\varphi}$ together with triangle-level curl potentials $\widehat{\psi}$. Because the LFP grid is fully triangulated, the complex has trivial first homology and the harmonic subspace is zero-dimensional, just as in the validation experiment of Section~\ref{ssec:simu_recovery}; on this complex, the Hodge decomposition therefore reduces to gradient and curl components, with the harmonic term identically zero by construction. The harmonic component remains an essential part of the framework for complexes with topological holes (Section~\ref{ssec:simu_harmonic}), but does not contribute here.

For all analyses below, the continuous LFP traces are segmented into non-overlapping 1\,s epochs, the lead-lag MI is computed within each epoch (lag $k_{\max} = 3$ samples), and the Hodge decomposition is applied to each resulting network. We compare 60 pre-pMCAo epochs against 60 post-pMCAo epochs per animal.

\subsection{Descriptive Reorganization of Information Flow}
\label{ssec:appl_descriptive}

\begin{figure}[H]
    \centering
    \includegraphics[width=0.9\linewidth]{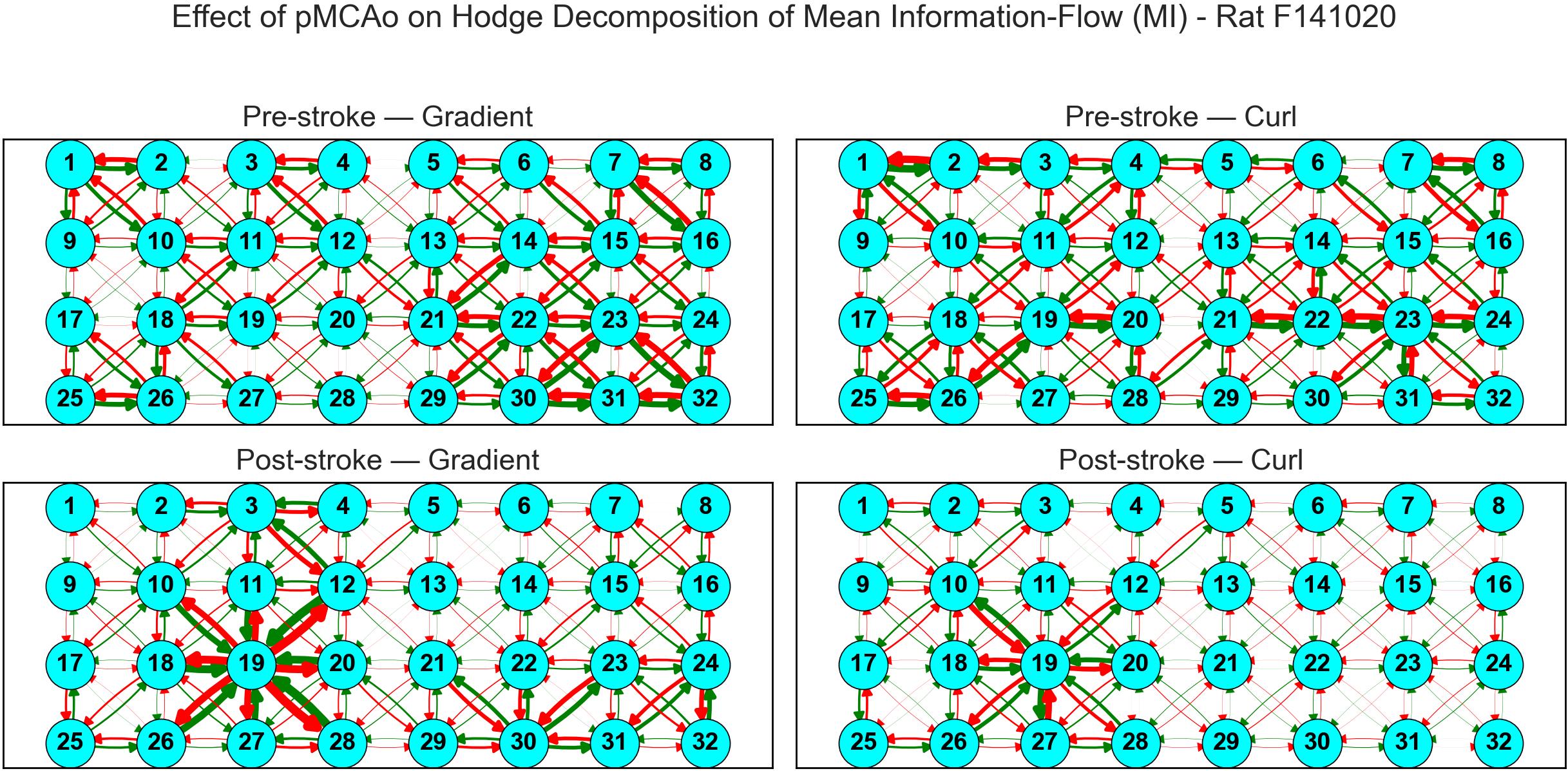}
    \caption{Hodge decomposition averaged across all epochs in each condition for Rat F141020. Top row: pre-pMCAo. Bottom row: post-pMCAo. Left: gradient component. Right: curl component. The harmonic component is identically zero on this fully triangulated grid (Section~\ref{ssec:simu_recovery}).}
    \label{fig:lfp_hd_components_stroke_disruption}
\end{figure}

We first present results for one representative animal (Rat F141020), and defer multi-rat results to Section~\ref{ssec:appl_testing} and Appendix~\ref{app:per_rat}. Figure~\ref{fig:lfp_hd_components_stroke_disruption} shows the mean gradient and curl components computed across pre- and post-pMCAo epochs. The gradient component (left panels) maps the dominant feed-forward pathways from driver to driven sites; the curl component (right panels) captures triangle-level recurrent loops among neighboring electrodes, illuminating how local integration is reconfigured around the MCA territory. As noted above, on this fully triangulated complex the harmonic component is identically zero by construction, so it is not displayed.

Even at this descriptive level the reorganization is substantial. Nodes 19, 32, and several superficial-layer sites stand out as the most reconfigured. Post-stroke, node 19 develops sink-like behavior, with surrounding nodes preferentially directing flow toward it, while node 32 (at the deepest, lateral edge of the array) becomes a stronger source. The gradient and curl panels both show visible changes from pre- to post-stroke (Fig.~\ref{fig:lfp_hd_components_stroke_disruption}), which we quantify formally in Section~\ref{ssec:appl_testing}.

\begin{figure}[H]
    \centering
    \includegraphics[width=0.9\linewidth]{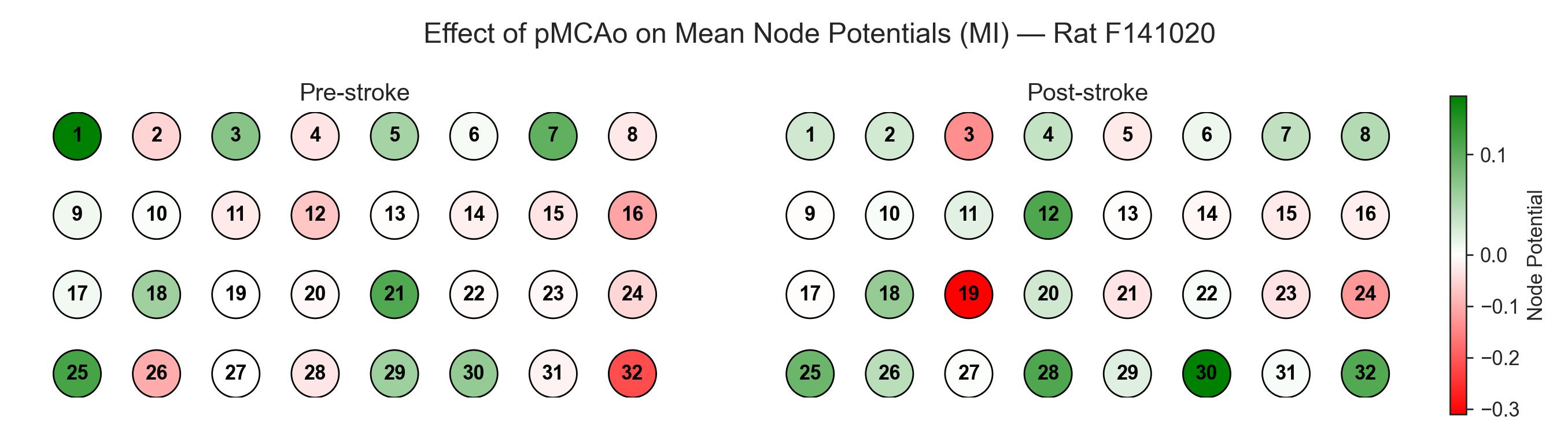}
    \caption{Node potentials averaged over 5\,min pre-pMCAo (left) and 5\,min post-pMCAo (right) for Rat F141020. Higher potential indicates a more source-like position in the feed-forward hierarchy.}
    \label{fig:lfp_hd_node_potential_stroke_disruption}
\end{figure}

The node-potential maps in Fig.~\ref{fig:lfp_hd_node_potential_stroke_disruption} summarize this hierarchical reorganization compactly. Nodes 12, 19, 30, and 32 show the largest shifts: node 19 drops in rank (more sink-like post-stroke) while node 32 rises (more source-like). Importantly these shifts are not randomly distributed across the array; they exhibit a vertical structure that, in the 4$\times$8 layered geometry of the implant, suggests a partial inversion of the canonical superficial-to-deep cortical hierarchy after ischemia.

\begin{figure}[H]
    \centering
    \includegraphics[width=0.9\linewidth]{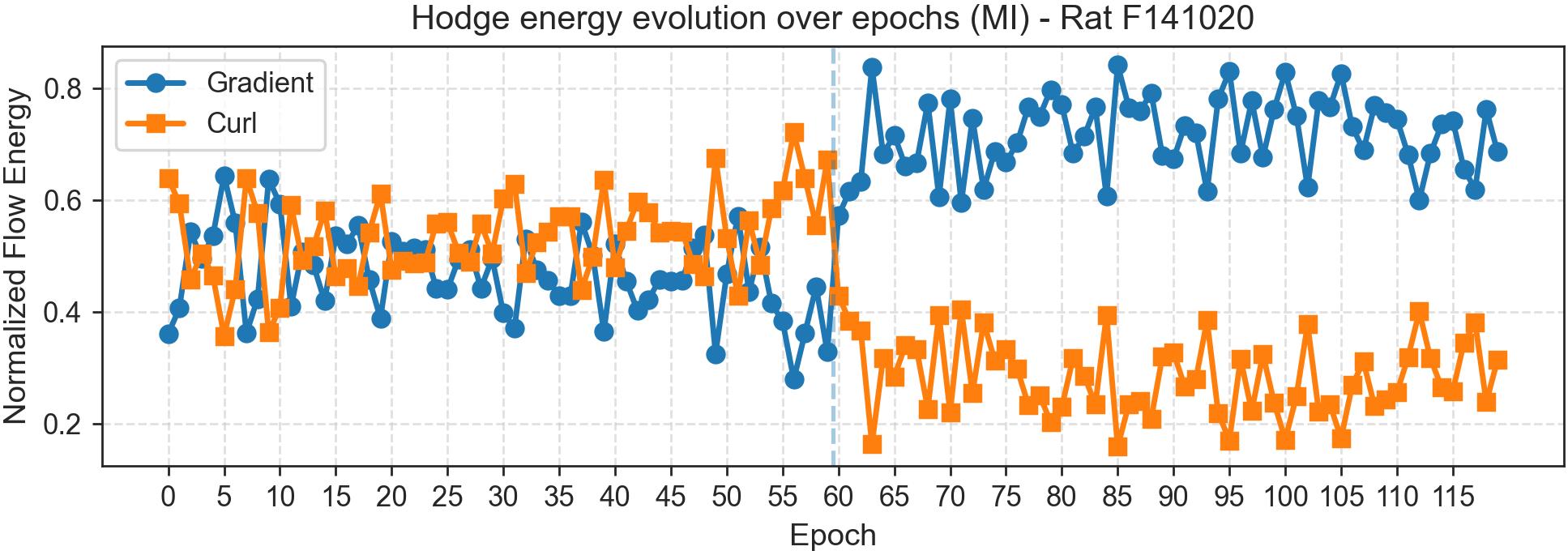}
    \caption{Normalized component energies over epochs for Rat F141020. For each 1\,s epoch we compute $\|-\delta_0\widehat{\varphi}\|_F^2 / \|W_a\|_F^2$ (Gradient, blue) and the analogous Curl ratio (orange). The harmonic component is identically zero on this fully triangulated grid (Section~\ref{ssec:simu_recovery}). The first 60 epochs precede pMCAo; the last 60 follow it.}
    \label{fig:lfp_hd_energy_evolution}
\end{figure}

Per-epoch changes in these components vary in magnitude across time and are often hard to detect by inspection alone. The energy trajectories in Fig.~\ref{fig:lfp_hd_energy_evolution} provide a global summary: for Rat F141020, gradient energy increases post-occlusion at the expense of the curl component, consistent with a relative loss of integrative feedback and a shift toward more strictly source-driven propagation. Other animals show qualitatively similar patterns of varying strength; one shows essentially no change. We document this heterogeneity in Appendix~\ref{app:per_rat}.

\subsection{Identifying Significantly Affected Nodes and Triangles}
\label{ssec:appl_testing}

Descriptive comparisons are informative but do not, by themselves, separate genuine post-stroke reorganization from epoch-to-epoch fluctuations in the recording. To address this we apply the block-permutation procedure of Section~\ref{ssec:testing} to identify nodes and triangular motifs whose Hodge potentials differ significantly between conditions, while properly controlling for the many simultaneous comparisons. We use $B = 5{,}000$ permutations, block length $6$ epochs, two-sided test statistics, and Benjamini--Hochberg FDR control at $\alpha = 0.05$.

\begin{figure}[H]
    \centering
    \includegraphics[width=0.95\linewidth]{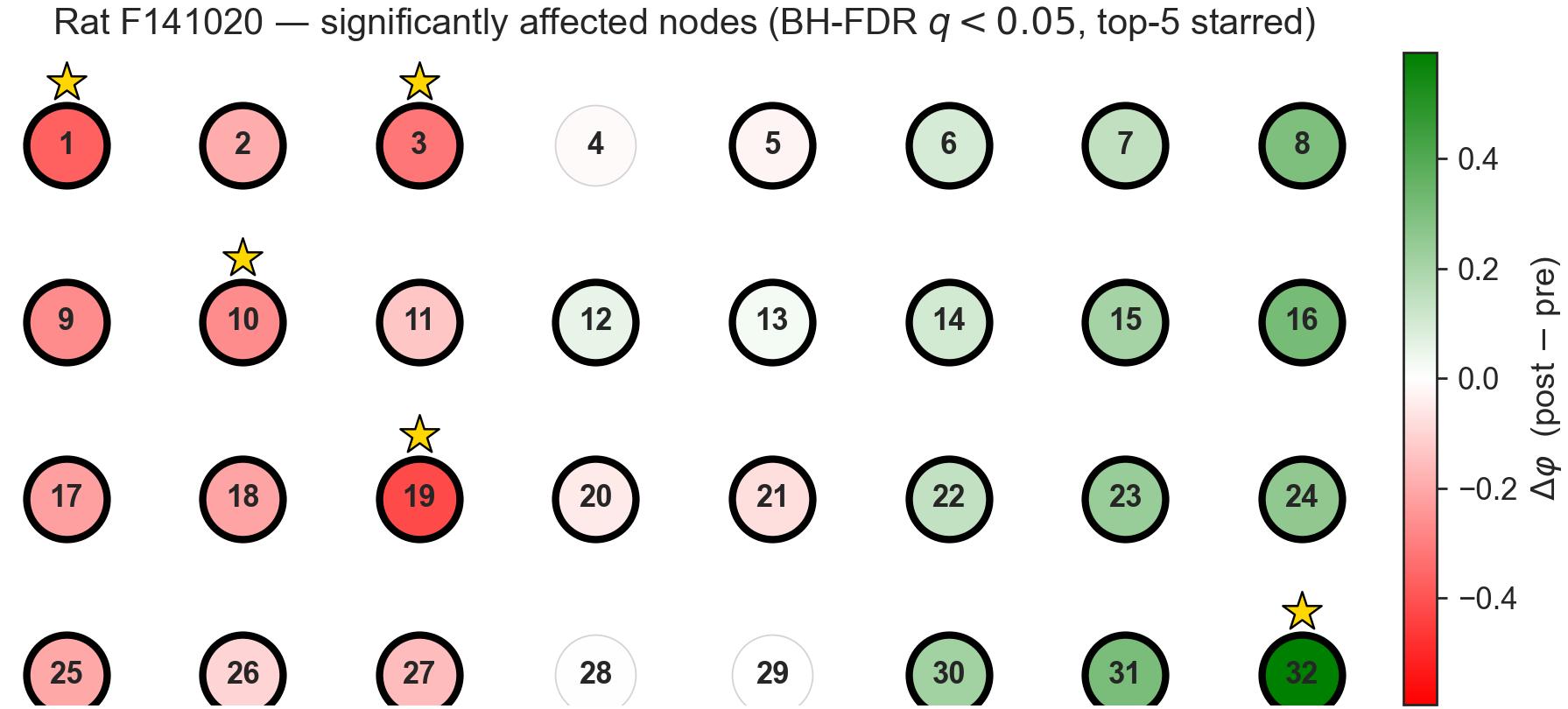}
    \caption{Significantly affected nodes for Rat F141020 (BH-FDR $q < 0.05$). Color encodes $\Delta\widehat\varphi$ (post $-$ pre); red indicates a drop in rank, green an increase. BH-significant nodes have black borders; the top five by $|T|$ are marked with a gold star.}
    \label{fig:appl_sig_nodes}
\end{figure}

\textbf{Node-level results.}
For Rat F141020, $29$ of $32$ nodes show statistically significant changes in $\widehat\varphi$ after BH correction. This is a high rate but it is consistent with the global character of the pMCAo lesion: ischemia in this preparation perturbs essentially the entire recorded territory, and our test honestly reports this. Because so many nodes pass the threshold, we rank the significant nodes by effect size $|T_p^{\varphi}|$ and focus on the five most strongly affected; the spatial pattern of the full set is shown in Fig.~\ref{fig:appl_sig_nodes}.

The top five nodes by effect size are 19 ($\Delta\widehat\varphi = -0.42$, layer V), 3 ($-0.32$, layer II/III), 10 ($-0.26$, layer IV), 1 ($-0.37$, layer II/III), and 32 ($+0.59$, layer Vb at the lateral edge). Two spatial patterns are visible in Fig.~\ref{fig:appl_sig_nodes}. Along the depth axis, superficial-layer sites (rows 1 and 2) shift toward sink-like roles while deep-layer sites (rows 3 and 4), and node 32 in particular, shift toward source-like roles, a pattern suggestive of a partial inversion of the canonical superficial-to-deep cortical hierarchy after ischemia. Along the horizontal axis, nodes on the right side of the array (columns 6--8) gain potential while nodes on the left side (columns 1--3) lose it. Because higher potential corresponds to a more source-like position in the feed-forward hierarchy, this means that post-stroke the right-side sites become net drivers and the left-side sites become net receivers, establishing a right-to-left feed-forward gradient along the cortical surface that was weaker or absent at baseline. The eight horizontal positions sample different locations within the MCA territory at $0.65\,\mathrm{mm}$ spacing, so this asymmetry may reflect differential proximity to the ischemic core, which does not affect all horizontal positions equally. Both the depth-axis and horizontal-axis observations are exploratory and based on a single animal; establishing either as a population-level effect would require a formal group-level test across animals. We emphasize that both interpretations are afforded by the geometry of the recording array; they are not built into the method itself.

\begin{figure}[H]
    \centering
    \includegraphics[width=0.95\linewidth]{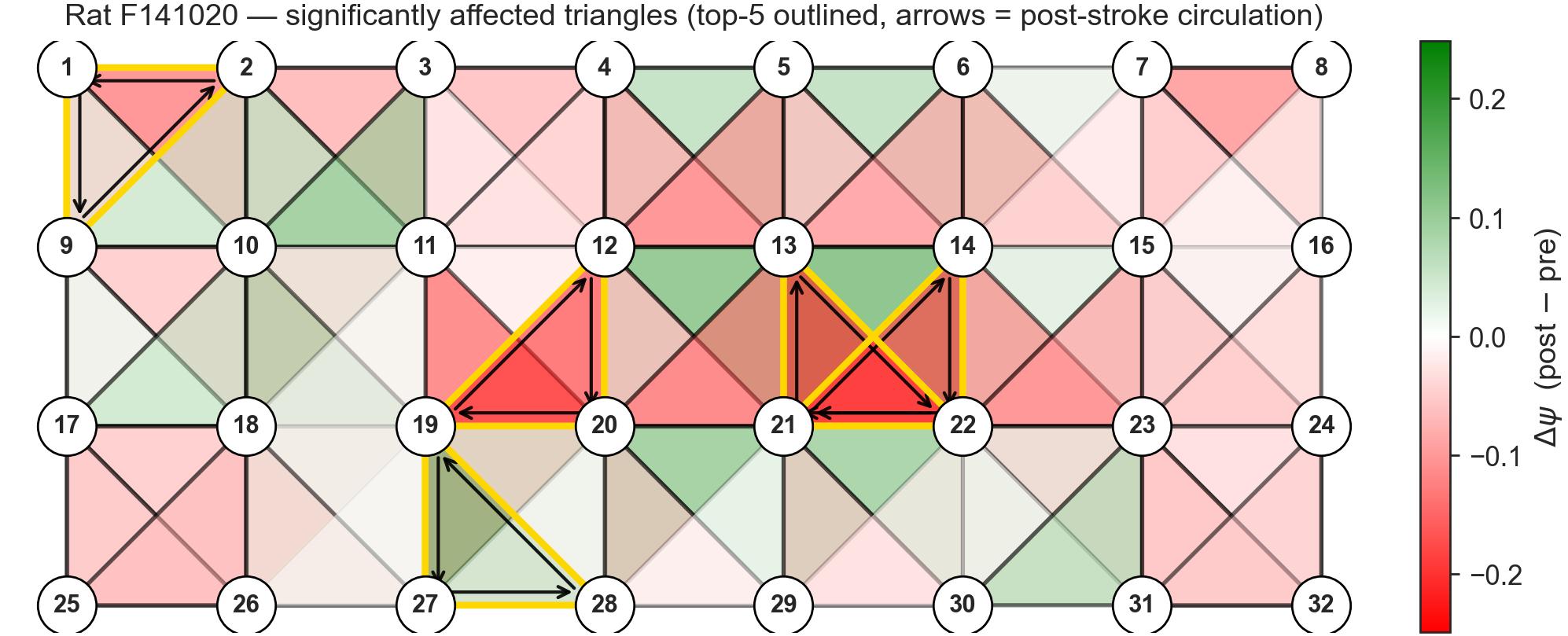}
    \caption{Significantly affected triangles for Rat F141020 (BH-FDR $q < 0.05$). Color encodes $\Delta\widehat\psi$; BH-significant patches have black outlines. The top five by $|T|$ are highlighted in gold, with arrows indicating post-stroke circulation direction.}
    \label{fig:appl_sig_triangles}
\end{figure}

\textbf{Triangle-level results.}
Testing the curl potentials $\widehat\psi_\tau$ flags $66$ of $84$ triangles as significant in Rat F141020 (BH-FDR $q < 0.05$). The interpretation of $\widehat\psi_\tau$ is more delicate than $\widehat\varphi_p$: under our convention $(p, q, r)$ with $p < q < r$, the sign of $\widehat\psi_\tau$ encodes the direction of circulation around the triangle: a positive value corresponds to flow $p \to q \to r \to p$ and a negative value to the reversed cycle $p \to r \to q \to p$. A significant change in $\widehat\psi_\tau$ between conditions therefore signals that either the magnitude or the rotation sense of a local recurrent loop has been reorganized. Figure~\ref{fig:appl_sig_triangles} shows the spatial distribution; for the five top-ranked triangles we overlay arrows indicating the post-stroke circulation direction.

The top five triangles by effect size are $\tau_{19,27,28}$ ($\Delta\widehat\psi = +0.16$), $\tau_{12,19,20}$ ($-0.23$), $\tau_{13,21,22}$ ($-0.25$), $\tau_{1,2,9}$ ($-0.11$), and $\tau_{14,21,22}$ ($-0.23$). Three of the five span layers IV and V, the canonical interface between thalamic input and intracortical output, and all three show negative $\Delta\widehat\psi$, indicating that triangle-level recurrent circulation across this interface changes magnitude or sign after stroke. The exception is $\tau_{19,27,28}$, located entirely within deep layers V/Vb, where a new recurrent loop emerges. Read together with the node-level results, this is consistent with the depth-axis reorganization described above. These are observations on a single animal in an exploratory analysis; they illustrate the kind of mechanistic statement the framework supports but do not establish a population-level claim.

\textbf{Summary of most significantly affected units.}
Table~\ref{tab:sig_summary} compiles the most strongly affected nodes and triangles, with their $\Delta(\cdot)$, BH-adjusted $q$-values, and brief interpretation.

\begin{table}[H]
\centering
\caption{Top five most strongly affected nodes and triangles in Rat F141020 (all $q < 0.05$, $B = 5{,}000$ block permutations). $\Delta\widehat\varphi$, $\Delta\widehat\psi$: post-minus-pre changes in mean potentials. Ordering is by $|T|$.}
\label{tab:sig_summary}
\begin{tabular}{lcccl}
\toprule
Unit & Type & $\Delta(\cdot)$ & $q$-value & Interpretation \\
\midrule
Node 19              & Node     & $-0.42$ & $<0.001$ & Layer V: becomes dominant sink \\
Node 3               & Node     & $-0.32$ & $<0.001$ & Layer II/III: rank drops, more sink-like \\
Node 10              & Node     & $-0.26$ & $<0.001$ & Layer IV: rank drops, more sink-like \\
Node 1               & Node     & $-0.37$ & $<0.001$ & Layer II/III: rank drops, more sink-like \\
Node 32              & Node     & $+0.59$ & $<0.001$ & Layer Vb (lateral edge): becomes dominant source \\
\midrule
$\tau_{19,27,28}$    & Triangle & $+0.16$ & $0.001$ & Deep layers V/Vb: new recurrent loop emerges \\
$\tau_{12,19,20}$    & Triangle & $-0.23$ & $0.001$ & Layer IV$\,\leftrightarrow\,$V feedback weakens/reverses \\
$\tau_{13,21,22}$    & Triangle & $-0.25$ & $0.001$ & Layer IV$\,\leftrightarrow\,$V feedback weakens/reverses \\
$\tau_{1,2,9}$       & Triangle & $-0.11$ & $0.001$ & Superficial II/III$\,\to\,$IV feedback weakens \\
$\tau_{14,21,22}$    & Triangle & $-0.23$ & $0.001$ & Layer IV$\,\leftrightarrow\,$V feedback weakens/reverses \\
\bottomrule
\end{tabular}
\end{table}

These results illustrate concretely why we view TECM as a useful tool. A coherence-based analysis applied to the same recordings would flag essentially every pair of channels involving node 19 as ``changed,'' without telling us whether the change reflects a shift in node 19's hierarchical position (gradient layer) or a reorganization of the recurrent circuits it participates in (curl layer). TECM disentangles these two phenomena, assigns statistical significance to each, and, thanks to the geometry of the implant, ties them to interpretable laminar structures within S1. The combination of a finer topological vocabulary and a principled multiple-testing procedure is what turns the qualitative observation ``node 19 looks different'' into the more specific statement that node 19's hierarchical role and its participation in layer IV$\,\leftrightarrow\,$V feedback both change after stroke.

\textbf{Robustness across animals.}
Applying the same testing pipeline to the three remaining animals confirms that the F141020 pattern is not an artifact of one subject but is also not universally present. Rats F150326 and F150410 show widespread significant reorganization at both topological scales (19 and 16 of 32 nodes; 49 and 34 of 84 triangles, respectively), with affected nodes organized along the depth axis in a manner broadly consistent with the F141020 pattern. In contrast, Rat F160406 shows no BH-significant nodes or triangles. We interpret this as a useful negative control: when reorganization is genuinely absent or small relative to recording variability, the test does not produce false positives. Detailed per-rat figures and summary statistics are reported in Appendix~\ref{app:per_rat}.

\section{Conclusion}
\label{sec:conc}

This paper introduces Topological Effective Connectivity Modeling (TECM), a framework that couples information-theoretic lead-lag dependence with the discrete Hodge decomposition to characterize directed information flow in networks where feedback is intrinsic. TECM splits edge flow into a gradient component supporting a node ordering, a curl component capturing triangle-level feedback, and a harmonic component capturing longer-range cyclic structure, and uses a permutation-based testing layer with FDR control to identify nodes and triangular motifs that change between conditions. The framework matters because traditional synchrony measures conflate hierarchical drive with recurrent feedback without distinguishing which region steers propagation from which triangle-level circuits amplify or gate it. TECM makes this distinction explicit and testable, and treats topological objects as principled units of inference, reducing the multiple-testing burden of edge-by-edge analyses while retaining a clean physical interpretation.

Applied to LFP recordings in a pMCAo rodent model, TECM reveals stroke-induced reorganization in both hierarchical drive and recurrent circuitry near the infarct territory; in three of four animals, formal testing flags spatial patterns organized along the depth axis of the recording array in a manner suggestive of laminar reorganization, a finding that needs confirmation at the group level. Three extensions stand out. First, group-level modeling would let us formally test the laminar reorganization picture across animals. Second, richer skeletons derived from anatomical connectivity or known long-range projections such as interhemispheric or thalamocortical pathways would supply nontrivial first homology and engage the harmonic component; the potential landscape is also a natural object to interrogate for traveling-wave phenomena. Third, coupling the present framework with formal interventional designs, for instance targeted stimulation experiments, would move the analysis from effective connectivity toward causal inference in the interventional sense, a direction we deliberately leave open in the current work.

\newpage
\appendix
\section{Per-rat results}
\label{app:per_rat}

The main paper focuses on Rat F141020 for the descriptive and testing analyses. Here we report the analogous spatial maps for the remaining three animals (F150326, F150410, F160406), together with a numerical summary across all four rats. The qualitative consistency across F141020, F150326, and F150410, contrasted with the negative-control behavior of F160406, supports the interpretation given in Section~\ref{ssec:appl_testing}.

\begin{table}[H]
\centering
\caption{Per-rat counts of BH-significant nodes and triangles ($\alpha = 0.05$), with the three most-affected nodes and triangles (by $|T|$) for each animal. Triangles are indexed by position in the sorted list of unique triangles.}
\label{tab:per_rat_summary}
\begin{tabular}{lrrll}
\toprule
Rat & \#sig nodes & \#sig triangles & Top-3 nodes & Top-3 triangle indices \\
\midrule
F141020 & 29 / 32 & 66 / 84 & [19, 3, 10]   & [66, 41, 46] \\
F150326 & 19 / 32 & 49 / 84 & [3, 25, 29]   & [46, 54, 55] \\
F150410 & 16 / 32 & 34 / 84 & [29, 32, 22]  & [76, 70, 16] \\
F160406 & \phantom{0}0 / 32 & \phantom{0}0 / 84 & [32, 19, 1] & [41, 83, 13] \\
\bottomrule
\end{tabular}
\end{table}

Three of the four animals (F141020, F150326, F150410) show widespread BH-significant reorganization at both topological scales. The top-ranked nodes vary across animals but share the property of lying along the depth dimension of the array, consistent with the laminar reorganization picture argued in the main text. Rat F160406 shows no BH-significant unit at either scale and serves as a useful negative control: the unranked top-3 lists for this animal still reflect the largest effect sizes observed, but their $|T|$ values are too small to clear the BH threshold.

\begin{figure}[H]
    \centering
    \begin{tabular}{cc}
    \includegraphics[width=0.46\linewidth]{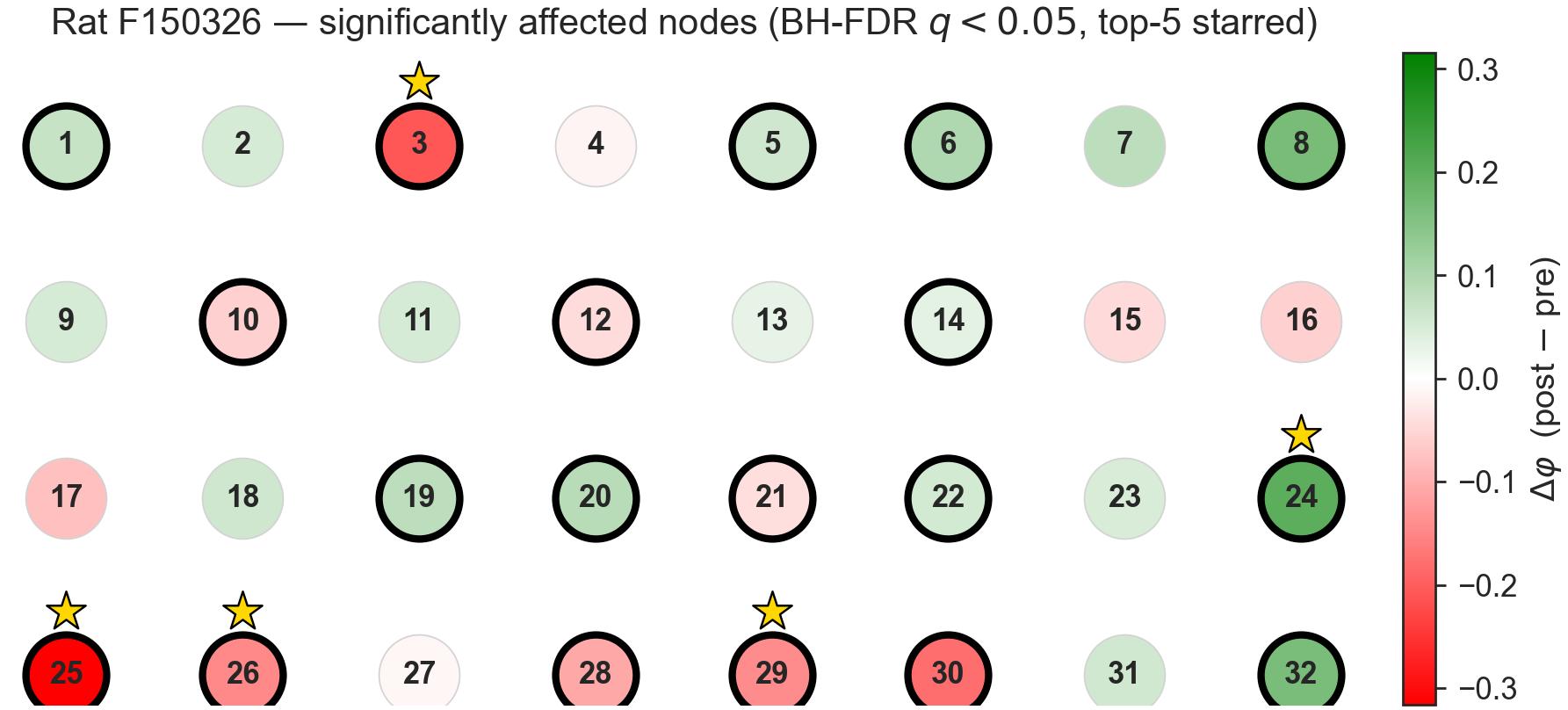} &
    \includegraphics[width=0.46\linewidth]{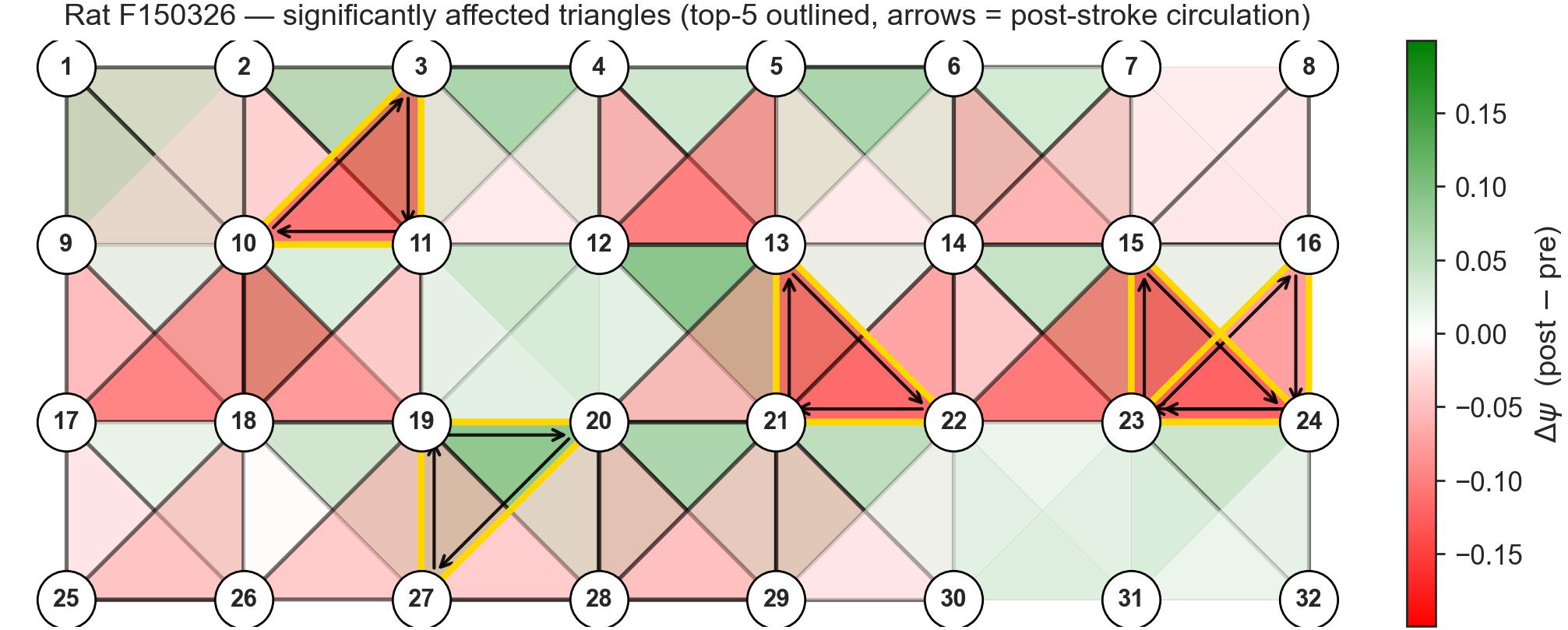} \\
    \multicolumn{2}{c}{\small (a) Rat F150326} \\[6pt]
    \includegraphics[width=0.46\linewidth]{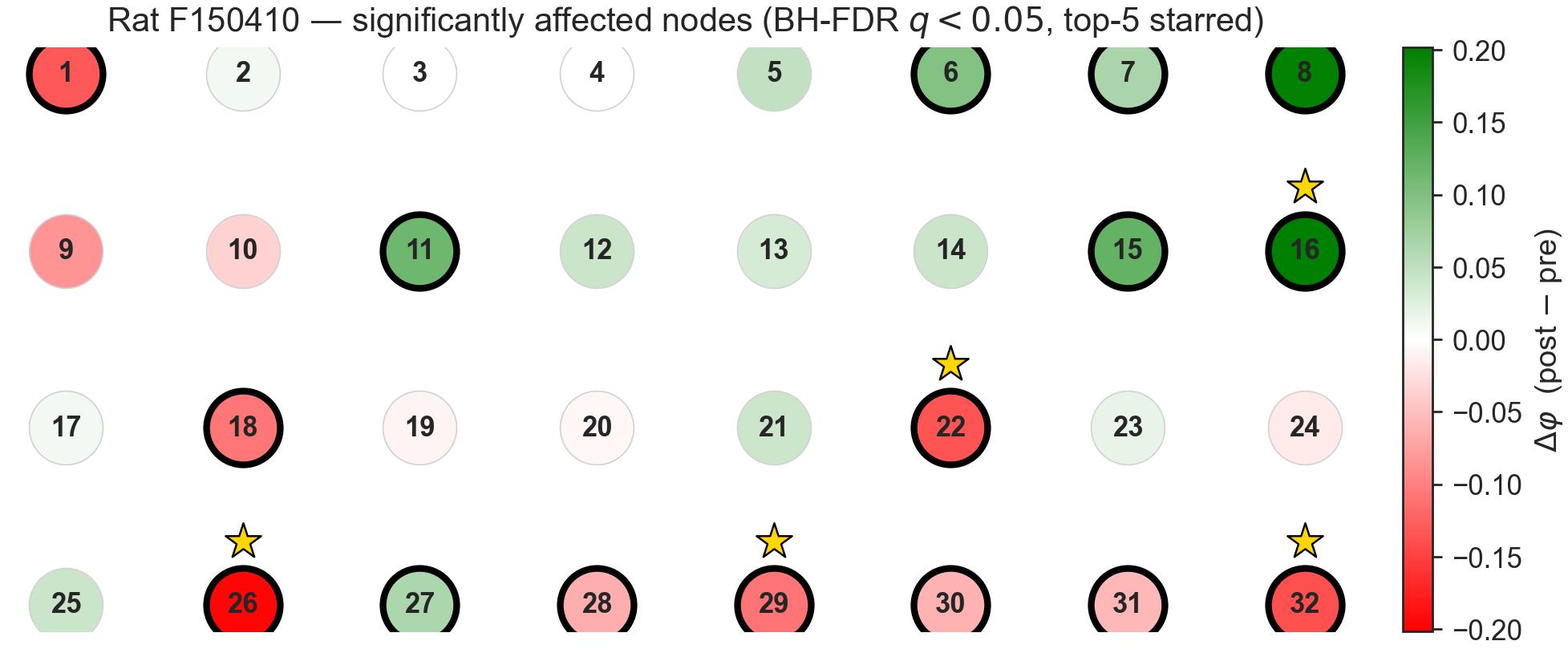} &
    \includegraphics[width=0.46\linewidth]{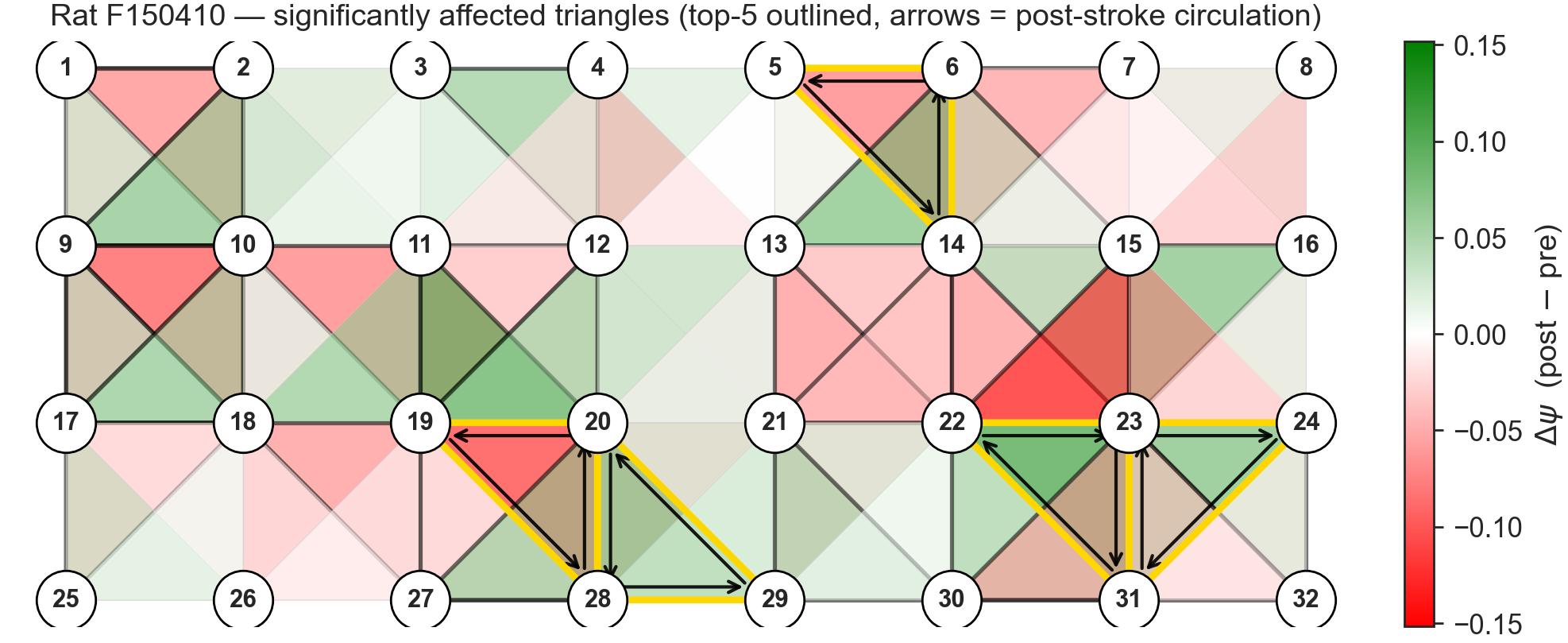} \\
    \multicolumn{2}{c}{\small (b) Rat F150410} \\[6pt]
    \includegraphics[width=0.46\linewidth]{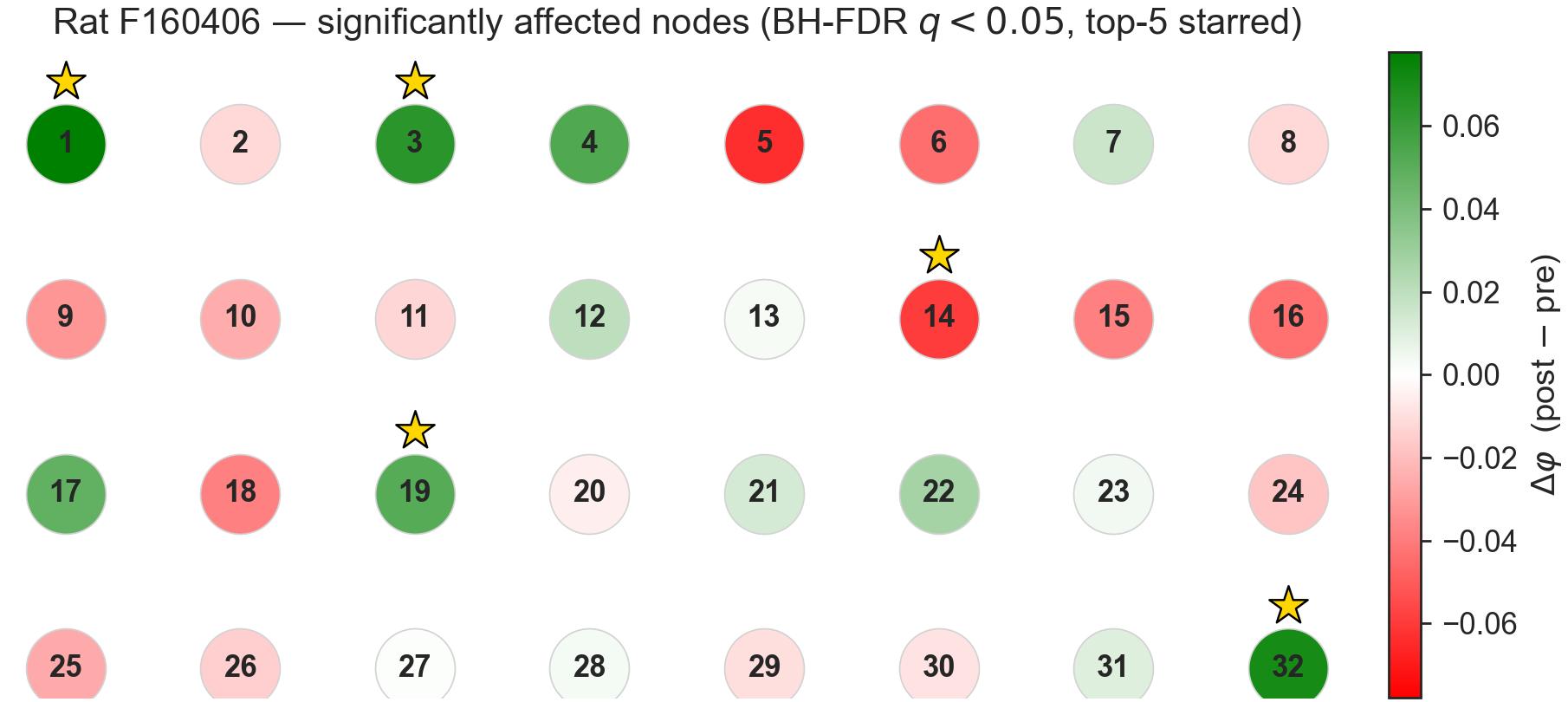} &
    \includegraphics[width=0.46\linewidth]{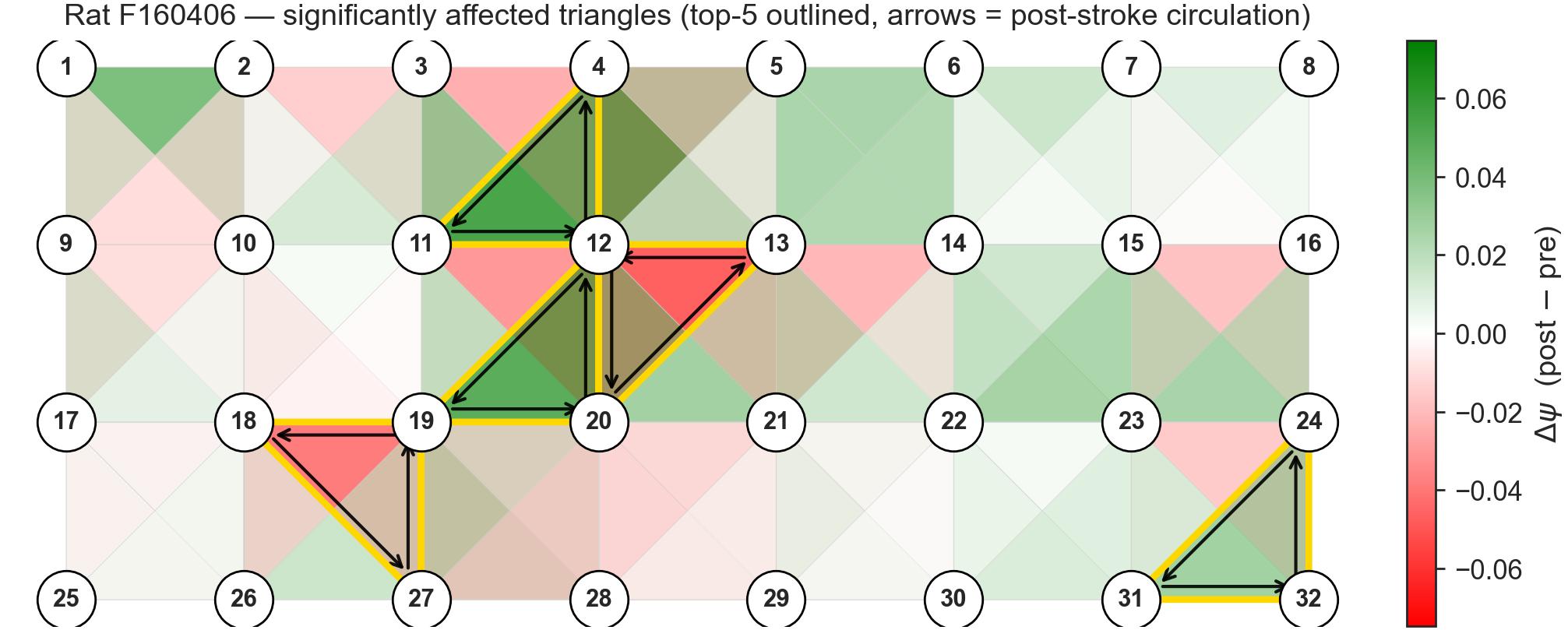} \\
    \multicolumn{2}{c}{\small (c) Rat F160406} \\
    \end{tabular}
    \caption{Per-rat spatial maps of significantly affected nodes (left) and triangles (right) for the three animals not shown in the main text. Conventions follow Figs.~\ref{fig:appl_sig_nodes} and \ref{fig:appl_sig_triangles}.}
    \label{fig:per_rat_maps}
\end{figure}






\newpage
\bibliographystyle{chicago}
\bibliography{references}

@article{BRAIN_NETWORKS_1,
    author    = {Friston, Karl J.},
    title     = {Functional and effective connectivity: a review},
    journal   = {Brain Connectivity},
    volume    = {1},
    number    = {1},
    pages     = {13--36},
    year      = {2011},
    publisher = {Mary Ann Liebert, Inc.},
    doi       = {10.1089/brain.2011.0008}
}

@book{BRAIN_NETWORKS_2,
    author    = {Fornito, Alex and Zalesky, Andrew and Bullmore, Edward},
    title     = {Fundamentals of Brain Network Analysis},
    year      = {2016},
    publisher = {Academic Press},
    address   = {San Diego, CA, USA},
    isbn      = {9780124079083}
}

@article{GRAPH_MODELING_COMPLEX_NETWORK,
    author  = {Stam, Cornelis J. and Reijneveld, Jaap C.},
    title   = {Graph theoretical analysis of complex networks in the brain},
    journal = {Nonlinear Biomedical Physics},
    year    = {2007},
    volume  = {1},
    number  = {1},
    pages   = {3},
    doi     = {10.1186/1753-4631-1-3}
}

@article{GRAPH_MODELING_HUMAN_BRAIN,
    author  = {Bassett, Danielle S. and Bullmore, Edward T.},
    title   = {Human brain networks in health and disease},
    journal = {Current Opinion in Neurology},
    volume  = {22},
    number  = {4},
    pages   = {340--347},
    year    = {2009},
    doi     = {10.1097/WCO.0b013e32832d93dd}
}

@article{GRAPH_MODELING_BRAIN_CONNECTIVITY,
    author  = {He, Yong and Evans, Alan},
    title   = {Graph theoretical modeling of brain connectivity},
    journal = {Current Opinion in Neurology},
    year    = {2010},
    volume  = {23},
    number  = {4},
    pages   = {341--350},
    doi     = {10.1097/WCO.0b013e32833aa567}
}

@article{HOI_1,
    author    = {Yu, Shan and Yang, Hongdian and Nakahara, Hiroyuki and Santos, Gustavo S. and Nikoli{\'c}, Danko and Plenz, Dietmar},
    title     = {Higher-order interactions characterized in cortical activity},
    journal   = {Journal of Neuroscience},
    volume    = {31},
    number    = {48},
    pages     = {17514--17526},
    year      = {2011},
    publisher = {Society for Neuroscience},
    doi       = {10.1523/JNEUROSCI.3127-11.2011}
}

@article{HOI_2,
    author  = {Herzog, Rub\'{e}n and Rosas, Fernando E. and Whelan, Robert and Fittipaldi, Sol and Santamaria-Garcia, Hernando and Cruzat, Josephine and Birba, Agustina and Moguilner, Sebastian and Tagliazucchi, Enzo and Prado, Pavel and Ibanez, Agustin},
    title   = {Genuine high-order interactions in brain networks and neurodegeneration},
    journal = {Neurobiology of Disease},
    volume  = {175},
    pages   = {105918},
    year    = {2022},
    issn    = {0969-9961},
    doi     = {10.1016/j.nbd.2022.105918}
}

@article{HOI_3,
    author  = {Santoro, Andrea and Battiston, Federico and Lucas, Maxime and Petri, Giovanni and Amico, Enrico},
    title   = {Higher-order connectomics of human brain function reveals local topological signatures of task decoding, individual identification, and behavior},
    journal = {Nature Communications},
    year    = {2024},
    volume  = {15},
    number  = {1},
    pages   = {1--12},
    doi     = {10.1038/s41467-024-54472-y}
}

@article{TDA_BRAIN,
    author  = {Lee, Hyekyoung and Kang, Hyejin and Chung, Moo K. and Kim, Bung-Nyun and Lee, Dong Soo},
    title   = {Persistent brain network homology from the perspective of dendrogram},
    journal = {IEEE Transactions on Medical Imaging},
    year    = {2012},
    volume  = {31},
    number  = {12},
    pages   = {2267--2277},
    doi     = {10.1109/TMI.2012.2219590}
}

@article{TDA_BRAIN_ARTERY,
    author  = {Bendich, Paul and Marron, J. S. and Miller, Ezra and Pieloch, Alex and Skwerer, Sean},
    title   = {Persistent homology analysis of brain artery trees},
    journal = {The Annals of Applied Statistics},
    year    = {2016},
    volume  = {10},
    number  = {1},
    pages   = {198--218},
    doi     = {10.1214/15-AOAS886}
}

@article{IT_1,
    author  = {Gatica, Marilyn and Cofr\'{e}, Rodrigo and Mediano, Pedro A.M. and Rosas, Fernando E. and Orio, Patricio and Diez, Ibai and Swinnen, Stephan P. and Cortes, Jesus M.},
    title   = {High-order interdependencies in the aging brain},
    journal = {Brain Connectivity},
    volume  = {11},
    number  = {9},
    pages   = {734--744},
    year    = {2021},
    doi     = {10.1089/brain.2020.0982}
}

@article{IT_2,
    author  = {Varley, Thomas F. and Mediano, Pedro A. M. and Patania, Alice and Bongard, Josh},
    title   = {The topology of synergy: Linking topological and information-theoretic approaches to higher-order interactions in complex systems},
    journal = {PLOS Computational Biology},
    volume  = {21},
    number  = {11},
    pages   = {e1013649},
    year    = {2025},
    month   = {11},
    doi     = {10.1371/journal.pcbi.1013649},
    publisher = {Public Library of Science}
}

@article{TDA_BRAIN_PROS_AND_CONS,
    author  = {Caputi, Luigi and Pidnebesna, Anna and Hlinka, Jaroslav},
    title   = {Promises and pitfalls of topological data analysis for brain connectivity analysis},
    journal = {NeuroImage},
    volume  = {238},
    pages   = {118245},
    year    = {2021},
    doi     = {10.1016/j.neuroimage.2021.118245}
}

@article{METHODS_BCA,
    author  = {El-Yaagoubi, Anass B. and Aslan, Sipan and Gomawi, Farah and Redondo, Paolo V. and Roy, Sarbojit and Sultan, Malik S. and Talento, Mara S. and Tarrazona, Francine T. and Wu, Haibo and Cooper, Keiland W. and Fortin, Norbert J. and Ombao, Hernando},
    title   = {Methods for Brain Connectivity Analysis with Applications to Rat Local Field Potential Recordings},
    journal = {Entropy},
    year    = {2025},
    volume  = {27},
    number  = {4},
    pages   = {328},
    doi     = {10.3390/e27040328}
}

@article{TDA_ORIENTED_1,
    author  = {Reimann, Michael W. and Nolte, Max and Scolamiero, Martina and Turner, Katharine and Perin, Rodrigo and Chindemi, Giuseppe and D{\l}otko, Pawe{\l} and Levi, Ran and Hess, Kathryn and Markram, Henry},
    title   = {Cliques of Neurons Bound into Cavities Provide a Missing Link between Structure and Function},
    journal = {Frontiers in Computational Neuroscience},
    volume  = {11},
    pages   = {48},
    year    = {2017},
    doi     = {10.3389/fncom.2017.00048}
}

@article{TDA_ORIENTED_2,
    author  = {Chowdhury, Samir and M{\'e}moli, Facundo},
    title   = {A functorial {D}owker theorem and persistent homology of asymmetric networks},
    journal = {Journal of Applied and Computational Topology},
    volume  = {2},
    number  = {1},
    pages   = {115--175},
    year    = {2018},
    doi     = {10.1007/s41468-018-0020-6},
    publisher = {Springer}
}

@incollection{TDA_ORIENTED_3,
    author    = {El-Yaagoubi, Anass B. and Ombao, Hernando},
    title     = {Topological Data Analysis for Directed Dependence Networks of Multivariate Time Series Data},
    booktitle = {Research Papers in Statistical Inference for Time Series and Related Models},
    editor    = {Liu, Yan and Hirukawa, Junichi and Kakizawa, Yoshihide},
    chapter   = {17},
    pages     = {403--417},
    publisher = {Springer},
    address   = {Singapore},
    year      = {2023},
    doi       = {10.1007/978-981-99-0803-5_17}
}

@inproceedings{TDA_ORIENTED_4,
    author    = {El-Yaagoubi, Anass B. and Chung, Moo K. and Ombao, Hernando},
    editor    = {Chen, Chao and Singh, Yash and Hu, Xiaoling},
    title     = {Topological Analysis of Seizure-Induced Changes in Brain Hierarchy Through Effective Connectivity},
    booktitle = {Topology- and Graph-Informed Imaging Informatics},
    series    = {Lecture Notes in Computer Science},
    volume    = {15239},
    year      = {2024},
    publisher = {Springer Nature Switzerland},
    pages     = {134--145},
    doi       = {10.1007/978-3-031-73967-5_13}
}

@article{TDA_INTRO_YAAGOUBI,
    author  = {El-Yaagoubi, Anass B. and Chung, Moo K. and Ombao, Hernando},
    title   = {Topological Data Analysis for Multivariate Time Series Data},
    journal = {Entropy},
    year    = {2023},
    volume  = {25},
    number  = {11},
    pages   = {1509},
    doi     = {10.3390/e25111509}
}

@inproceedings{CI_ASSUMPTIONS_1,
    author    = {Druzdzel, Marek J.},
    title     = {The role of assumptions in causal discovery},
    booktitle = {8th Workshop on Uncertainty Processing (WUPES-09)},
    pages     = {57--68},
    month     = {September},
    year      = {2009}
}

@article{CI_ASSUMPTIONS_2,
    author  = {Vonk, Maarten C. and Malekovic, Ninoslav and B{\"a}ck, Thomas and Kononova, Anna V.},
    title   = {Disentangling Causality: Assumptions in Causal Discovery and Inference},
    journal = {Artificial Intelligence Review},
    year    = {2023},
    volume  = {56},
    number  = {9},
    pages   = {10613--10649},
    doi     = {10.1007/s10462-023-10411-9}
}

@book{COVER_THOMAS_2006,
    author    = {Cover, Thomas M. and Thomas, Joy A.},
    title     = {Elements of Information Theory},
    edition   = {2},
    year      = {2006},
    publisher = {Wiley-Interscience},
    address   = {Hoboken, NJ},
    isbn      = {978-0471241959}
}

@article{KRASKOV_MI,
    author  = {Kraskov, Alexander and St\"{o}gbauer, Harald and Grassberger, Peter},
    title   = {Estimating mutual information},
    journal = {Physical Review E},
    volume  = {69},
    number  = {6},
    pages   = {066138},
    year    = {2004},
    doi     = {10.1103/PhysRevE.69.066138}
}

@article{HODGE_RANK,
    author  = {Jiang, Xiaoye and Lim, Lek-Heng and Yao, Yuan and Ye, Yinyu},
    title   = {Statistical ranking and combinatorial Hodge theory},
    journal = {Mathematical Programming},
    volume  = {127},
    number  = {1},
    pages   = {203--244},
    year    = {2011},
    doi     = {10.1007/s10107-010-0419-x}
}

@article{HODGE_DNA,
    author  = {Wei, Ronald Koh Joon and Wee, Junjie and Laurent, Valerie Evangelin and Xia, Kelin},
    title   = {Hodge theory-based biomolecular data analysis},
    journal = {Scientific Reports},
    year    = {2022},
    volume  = {12},
    number  = {1},
    pages   = {9699},
    doi     = {10.1038/s41598-022-12877-z}
}

@article{HODGE_GRAPHS,
    author  = {Lim, Lek-Heng},
    title   = {Hodge Laplacians on Graphs},
    journal = {SIAM Review},
    volume  = {62},
    number  = {3},
    pages   = {685--715},
    year    = {2020},
    doi     = {10.1137/18M1223101}
}

@article{BENJAMINI_HOCHBERG_1995,
    author  = {Benjamini, Yoav and Hochberg, Yosef},
    title   = {Controlling the false discovery rate: a practical and powerful approach to multiple testing},
    journal = {Journal of the Royal Statistical Society: Series B (Methodological)},
    volume  = {57},
    number  = {1},
    pages   = {289--300},
    year    = {1995},
    doi     = {10.1111/j.2517-6161.1995.tb02031.x}
}

@article{BENJAMINI_YEKUTIELI_2001,
    author  = {Benjamini, Yoav and Yekutieli, Daniel},
    title   = {The control of the false discovery rate in multiple testing under dependency},
    journal = {The Annals of Statistics},
    volume  = {29},
    number  = {4},
    pages   = {1165--1188},
    year    = {2001},
    doi     = {10.1214/aos/1013699998}
}

@article{NICHOLS_HOLMES_2002,
    author  = {Nichols, Thomas E. and Holmes, Andrew P.},
    title   = {Nonparametric permutation tests for functional neuroimaging: a primer with examples},
    journal = {Human Brain Mapping},
    volume  = {15},
    number  = {1},
    pages   = {1--25},
    year    = {2002},
    doi     = {10.1002/hbm.1058}
}

@article{POLITIS_ROMANO_1994,
    author  = {Politis, Dimitris N. and Romano, Joseph P.},
    title   = {The stationary bootstrap},
    journal = {Journal of the American Statistical Association},
    volume  = {89},
    number  = {428},
    pages   = {1303--1313},
    year    = {1994},
    doi     = {10.1080/01621459.1994.10476870}
}

@article{KUNSCH_1989,
    author  = {K{\"u}nsch, Hans R.},
    title   = {The jackknife and the bootstrap for general stationary observations},
    journal = {Annals of Statistics},
    volume  = {17},
    number  = {3},
    pages   = {1217--1241},
    year    = {1989},
    doi     = {10.1214/aos/1176347265}
}

@article{ISCHEMIC_STROKE_RESEARCH,
    author  = {Barthels, Derek and Das, Hiranmoy},
    title   = {Current advances in ischemic stroke research and therapies},
    journal = {Biochimica et Biophysica Acta (BBA) - Molecular Basis of Disease},
    volume  = {1866},
    number  = {4},
    pages   = {165260},
    year    = {2020},
    doi     = {10.1016/j.bbadis.2018.09.012}
}

@article{ISCHEMIC_STROKE_IMAGING_1,
    author  = {Smith, Aubrey George and Hill, Chris Rowland},
    title   = {Imaging assessment of acute ischaemic stroke: a review of radiological methods},
    journal = {British Journal of Radiology},
    year    = {2018},
    volume  = {91},
    number  = {1083},
    pages   = {20170573},
    doi     = {10.1259/bjr.20170573}
}

@article{ISCHEMIC_STROKE_IMAGING_2,
    author  = {Akbarzadeh, Mohammad Amin and Sanaie, Sarvin and Kuchaki Rafsanjani, Mahshid and Hosseini, Mohammad-Salar},
    title   = {Role of imaging in early diagnosis of acute ischemic stroke: a literature review},
    journal = {The Egyptian Journal of Neurology, Psychiatry and Neurosurgery},
    year    = {2021},
    volume  = {57},
    number  = {1},
    pages   = {175},
    doi     = {10.1186/s41983-021-00432-y}
}

@article{RAT_STROKE_EXPERIMENT_1,
    author  = {Wann, Ellen G. and Wodeyar, Anirudh and Srinivasan, Ramesh and Frostig, Ron D.},
    title   = {Rapid development of strong, persistent, spatiotemporally extensive cortical synchrony and underlying oscillations following acute MCA focal ischemia},
    journal = {Scientific Reports},
    year    = {2020},
    volume  = {10},
    number  = {1},
    pages   = {21441},
    doi     = {10.1038/s41598-020-78179-4}
}

@article{RAT_STROKE_EXPERIMENT_2,
    author  = {Rasheed, Waqas and Wodeyar, Anirudh and Srinivasan, Ramesh and Frostig, Ron D.},
    title   = {Sensory stimulation-based protection from impending stroke following {MCA} occlusion is correlated with desynchronization of widespread spontaneous local field potentials},
    journal = {Scientific Reports},
    year    = {2022},
    volume  = {12},
    number  = {1},
    pages   = {1744},
    doi     = {10.1038/s41598-022-05604-1}
}

@book{PEARL_CAUSALITY,
    author    = {Pearl, Judea},
    title     = {Causality: Models, Reasoning, and Inference},
    edition   = {2},
    year      = {2009},
    publisher = {Cambridge University Press},
    address   = {Cambridge, UK},
    isbn      = {978-0521895606}
}

@article{RUBIN_1974,
    author    = {Rubin, Donald B.},
    title     = {Estimating causal effects of treatments in randomized and nonrandomized studies},
    journal   = {Journal of Educational Psychology},
    volume    = {66},
    number    = {5},
    pages     = {688--701},
    year      = {1974},
    doi       = {10.1037/h0037350}
}

\end{document}